\immediate\write18{makeindex main.nlo -s nomencl.ist -o main.nls}
\documentclass[journal]{IEEEtran}

\usepackage{graphicx} 
\usepackage{amsmath} 
\usepackage{amssymb}
\usepackage{color}
\usepackage{epstopdf}  
\usepackage{cite}
\usepackage{algorithm}
\usepackage{multirow}
\usepackage{booktabs}
\usepackage{enumitem}
\usepackage{subfigure}
\usepackage{url}
\usepackage{nomencl}
\usepackage{ifthen}

\renewcommand{\nomgroup}[1]{%
	\item[\textbf{%
		\ifthenelse{\equal{#1}{M}}{Parameters}{}%
		\ifthenelse{\equal{#1}{N}}{Variables}{}
		\ifthenelse{\equal{#1}{I}}{Indexes}{}
		\ifthenelse{\equal{#1}{A}}{\hspace{-12pt} Superscripts}{}
	}]%
	
}

\makenomenclature

\def\N {\mathcal N}

\def\T{\mathcal T}

\def\K{\mathcal K}

\def\E{\mathcal E}
\def\jay{\mathrm{j}}

\title{Optimal Energy Scheduling and Sensitivity Analysis for Integrated Power-Water-Heat Systems}

\author{Sidun Fang, Chenxu Wang, Yashen Lin, and 
Changhong Zhao%
\thanks{This work was supported by the Hong Kong Research Grants Council though ECS Award 24210220 and the CUHK Faculty Startup Fund.}
\thanks{S. Fang, C. Wang, and C. Zhao are with the Department of Information Engineering, the Chinese University of Hong Kong, New Territories, Hong Kong SAR, China (Email: chzhao@ie.cuhk.edu.hk). Y. Lin is with the National Renewable Energy Laboratory, Golden, CO, USA.}
\thanks{\copyright 2021 IEEE. Personal use of this material is permitted. Permission from IEEE must be obtained for all other uses, in any current or future media, including reprinting/republishing this material for advertising or promotional purposes, creating new collective works, for resale or redistribution to servers or lists, or reuse of any copyrighted component of this work in other works.}
\thanks{This is the accepted version of an article to be published in IEEE Systems Journal. DOI: 10.1109/JSYST.2021.3127934}
}

\begin{document}

\maketitle

\begin{abstract}
The conventionally independent power, water, and heating networks are becoming more tightly connected, which motivates their joint optimal energy scheduling to improve the overall efficiency of an integrated energy system. However, such a joint optimization is known as a challenging problem with complex network constraints and couplings of electric, hydraulic, and thermal models that are nonlinear and nonconvex.
We formulate an optimal power-water-heat flow (OPWHF) problem and develop a computationally efficient heuristic to solve it. 
The proposed heuristic decomposes OPWHF into subproblems, which are iteratively solved via convex relaxation and convex-concave procedure.
Simulation results validate that the proposed framework can improve operational flexibility and social welfare of the integrated system, wherein the water and heating networks respond as virtual energy storage to time-varying energy prices and solar photovoltaic generation. Moreover, we perform sensitivity analysis to compare two modes of heating network control: by flow rate and by temperature. Our results reveal that the latter is more effective for heating networks with a wider space of pipeline parameters. 
\end{abstract}


\nomenclature[M]{\(r, x, z\)}{Line resistance, reactance and impedance}
\nomenclature[M]{\(\Delta t\)}{Duration of a time slot}
\nomenclature[M]{\(S\)}{Cross-sectional areas of tanks}
\nomenclature[M]{\(A,B,C\)}{Parameters of pumps}
\nomenclature[M]{\(c\)}{Specific heat capacity of water}
\nomenclature[M]{\(d\)}{Water demand}
\nomenclature[M]{\(\rho\)}{Mass density of water}
\nomenclature[M]{\(g\)}{Gravity constant}
\nomenclature[M]{\(\eta\)}{Pump efficiency}
\nomenclature[M]{\(F\)}{Friction parameter}
\nomenclature[M]{\(\xi\)}{Heat transfer coefficient}
\nomenclature[M]{\(\alpha\)}{Configuration of CHP operation}
\nomenclature[N]{\(P, Q\)}{Active and reactive power flows on lines}
\nomenclature[N]{\(p, q\)}{(with a node index subscript, e.g., $i$, $j$, $k$) Nodal active and reactive power injections}
\nomenclature[N]{\(q\)}{(with a line index subscript $ij$) Water flow rate}
\nomenclature[N]{\(v, \ell\)}{Squared nodal voltage and line current magnitudes} 
\nomenclature[N]{\(h\)}{Hydraulic head} 
\nomenclature[N]{\(H\)}{Heating power} 
\nomenclature[N]{\(\tau\)}{Water temperature} 
\nomenclature[A]{\(r\)}{Renewable energy sources}
\nomenclature[A]{\(\textnormal{gen}\)}{Generation nodes}
\nomenclature[A]{\(\textnormal{load}\)}{Load nodes}
\nomenclature[A]{\(\textnormal{CHP}\)}{Combined heat and power generators}
\nomenclature[A]{\(\textnormal{R,S}\)}{Return, Supply}
\printnomenclature

\section{Introduction}

Electricity, municipal water, fuels, and heat are conventionally transmitted in separate networks. 
To better utilize the flexibility of individual energy carriers to improve their efficiency and reduce emissions, different energy networks are more tightly integrated to form \textit{multi-energy systems}, or \textit{integrated energy systems} (IESs). 
Literature verified improvement in operational flexibility and economic benefits achieved by the joint energy scheduling of power-heat networks \cite{li2016DHoptLinear,li2016DHoptNonlinear}, power-water networks \cite{zamzam2018optimal,Oikonomou-2017}, and power-gas networks \cite{clegg2015integrated, yao2018stochastic}.
Such additional flexibility mainly comes from the gravitational potential energy stored in tanks \cite{zamzam2018optimal} and the thermal inertia of heating demand \cite{GU2017234, Lu2020inertia} and heating pipes \cite{mitridati2018power, li2016DHoptNonlinear}, which serve as energy storage to an IES. 

Compared to power networks, the water, fuel, and heating networks typically have slower dynamics that cannot be ignored even in hourly dispatch, including the hydraulic dynamics \cite{MSingh2020waterdynamics}, temperature propagation \cite{ARABKOOHSAR2019432}, and pressure losses \cite{jia2020convex}. 
To reduce computational burden, those dynamics can be approximated in discrete time as difference-algebraic equations, which contain bilinear or quadratic terms and are thus nonlinear and nonconvex. 
Convex relaxation \cite{jia2020convex} and piecewise linearization \cite{ding2019energy} are commonly exploited to approximately solve those equations. 

Specifically, in heating networks, two categories of assumptions have been made to simplify the bilinear model. The first assumes fixed water flow rates and controls water supply temperature to meet heating demand \cite{CHERTKOV201922}. This mode is widely adopted in China, Russia, some Nordic countries, et cetera. 
Based on the fixed-flow assumption, references \cite{Yang2019Laplace, PAN2017395, Li2019Energyhub, li2016DHoptLinear} reformulated optimal heat flow as linear or second-order cone (SOC) programs. 
The second category of assumptions simplifies the temperature propagation process in heating pipes. References \cite{antunes2008optimal, LIU20161238} assumed the temperature at junctions of pipes to be fixed.
In \cite{HUANG2019464}, the temperature of the maximal inlet water flow at a junction was used to approximate the temperature of all the outlet water flows. Recent work \cite{Convex-concaveHeating} exploited convex-concave method \cite{lipp2016variations} to reduce SOC relaxation error for joint power-gas-heat optimization. 
In water networks, to deal with bilinear terms of hydraulic heads and flow rates, reference \cite{fooladivanda2017energy} utilized exact SOC relaxation and \cite{zamzam2018optimal} took a successive convex inner approximation approach. 

Besides the physical model of an IES, its ownership is also a fundamental factor to consider.
There are generally two types of ownerships for IESs. In the first type, different energy carriers are managed by different companies who have their own benefits and may collaborate and/or compete on an integrated energy market \cite{YangCao2019Heatmarket, YueChen2019Heatmarket}. 
Reference \cite{dou2020bi} proposed a bi-level bidding retail market encompassing electricity, heat, and natural gas. 
Reference \cite{ge2021joint} proposed a region-user market mechanism for peer-to-peer transactions of integrated energy services.
Another type of IES is a multi-energy microgrid, e.g., in a factory \cite{huang2017planning}, an agriculture farm \cite{rincon2019contribution}, a residential community \cite{fang2020coordinated}, or a seaport \cite{fang2019toward}. Such an IES is usually operated by a single entity that optimizes the overall benefit of the included sectors (electricity, water, heating, etc.). There is still information privacy issue between different sectors when they are coordinated by the IES operator, which motivates the distributed solution process proposed in Section \ref{subsec:decomposed_solution}. Furthermore, the overall benefit of such an IES needs to be appropriately allocated between multiple sectors to provide incentives for joint operation, which is a topic worth further study beyond this paper. This paper considers the second type of IES containing three sectors coordinated by one operator: electric power, water, and heat.

Supplement to the literature above, this paper makes the following contributions:
\begin{enumerate}
\item To fully explore the potential of joint operations of multiple energy carriers, we propose a framework to co-optimize power, water, and heat flows subject to a relatively comprehensive set of network constraints. This is an advancement compared to prior studies \cite{li2016DHoptLinear,li2016DHoptNonlinear,zamzam2018optimal,Oikonomou-2017,Fang2021Continous} that each considered two kinds of energy flows and those ignoring network topology, e.g., \cite{karkhaneh2020risk}.

\item Different from the centralized solution methods, e.g., that in \cite{Convex-concaveHeating}, we decompose the optimal power-water-heat flow problem into subproblems that each require less computational effort and less information disclosed to a central operator. We apply a heuristic algorithm based on convex SOC relaxation and convex-concave procedure to iteratively and approximately solve the subproblems.

\item Through numerical simulations, we not only validate convergence and efficiency of the proposed framework, but also investigate the mechanism underlying the economic improvement made by the joint optimization. We also test the influences of uncertainties in solar generation and power, water, and heating demands on the performance of the proposed framework, and identify appropriate application scenarios of two different modes to control heating networks, i.e., by temperature and by flow rate. 
\end{enumerate}

The rest of this paper is organized as follows. Section \ref{sec:model} introduces a mathematical model of the integrated power, water, and heating network. Based on that model, Section \ref{sec:problem} formulates the optimal power-water-heat flow problem. To solve this problem, a heuristic method based on decomposed iterative convex approximation is presented in Section \ref{sec:solution}. Section \ref{sec:simulation} shows our simulation results to verify efficacy of the proposed framework, and Section \ref{sec:conclusion} concludes the paper.

\section{System Modeling}
\label{sec:model}

\begin{figure*}
	\includegraphics[width=1.4\columnwidth]{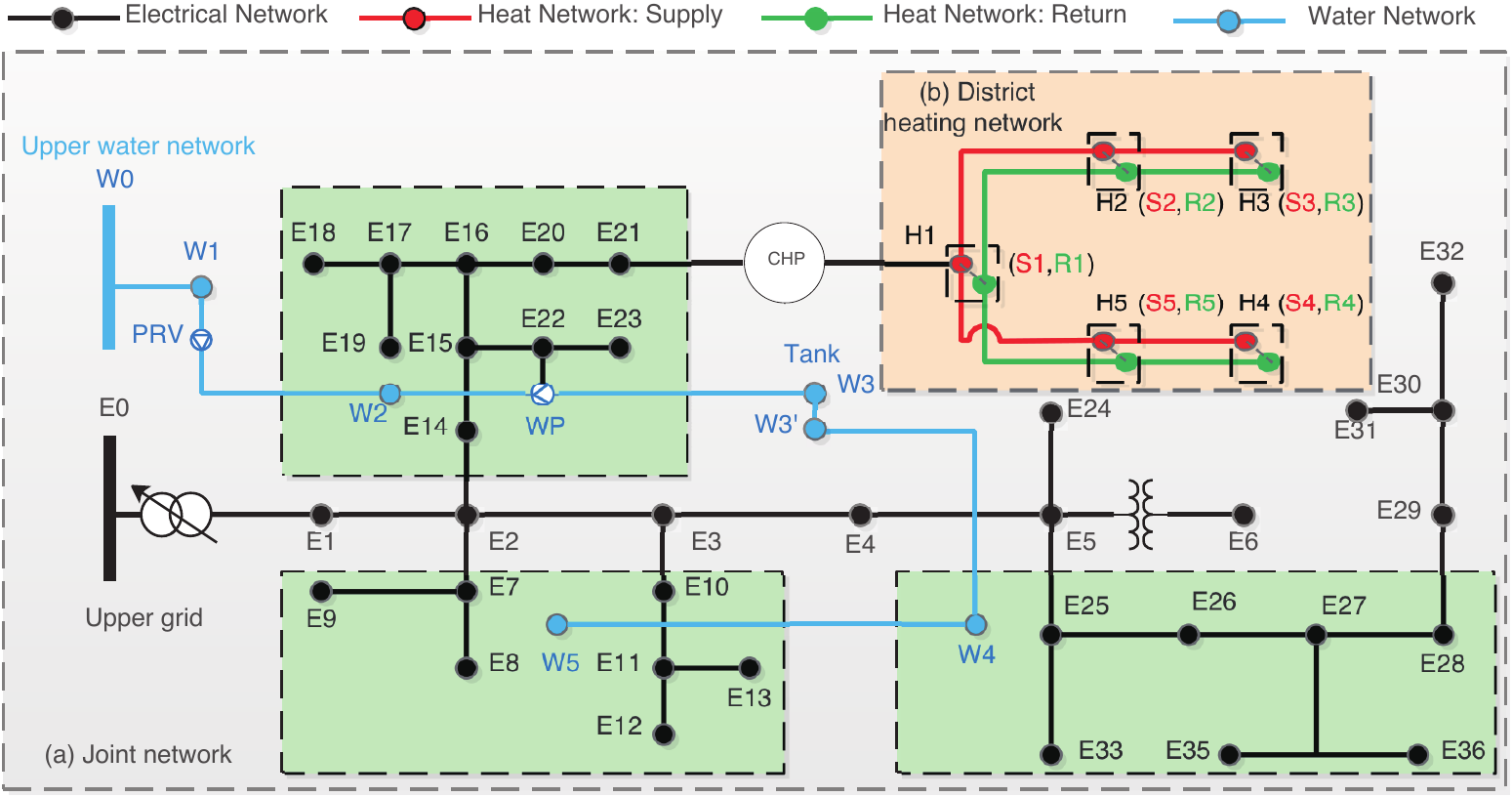}
	\centering
	\caption{A power-water-heat integrated system.}
	\label{fig:1}
\end{figure*}

Consider an integrated energy system comprising a power distribution network, a municipal water network, and a district heating network, as shown in Figure \ref{fig:1}. 
The power and water networks are connected by the electric pump at node E22 of the power network and node WP of the water network. The pump consumes electricity to elevate the hydraulic head and maintain the water flow. The relationship between the water flow rate, hydraulic head elevation, and electric power consumption of a pump is modeled later in \eqref{eq:MWN:pump-power}. 
The power and heating networks are coupled by the combined heat and power generator (CHP) at node E21 of the power network and node H1 of the heating network. The CHP produces electricity and heat simultaneously as modeled later in Section \ref{subsec:heat-model}.

Consider operating time horizon $\T :=\left\{1,2,\dots,|\T| \right\}$. A common example is $|\T| = 24$ with each $t \in \T$ representing an hour \cite{lahdelma2003CHP, sandou2005DHopt, li2016DHoptNonlinear}. Our main assumptions are:
\begin{enumerate}
\item We adopt the quasi-steady-state model of the heating network, which is suitable for hourly dispatch \cite{dancker2021improved}.
\item We simplify the water flow model by ignoring turbulence and laminar flows  \cite{zamzam2018optimal}.
\item The feasible operating region of a CHP is modeled as a convex polyhedron \cite{lahdelma2003CHP}. 
\end{enumerate}
We elaborate our model by parts in Sections \ref{subsec:power-model} to \ref{subsec:heat-model}. For convenience to look up, we list all the operational constraints in these sections in Table \ref{Table: Listofconstraints}. 

\begin{table}\centering
	\caption{List of operational constraints}\label{Table: Listofconstraints}
\begin{tabular}{lll}
	\hline
	Categories                       & Constraints               & Equations \\ \hline
	\multirow{3}{*}{Power network}   & Power flow equations      &     Eqn. (\ref{eq:PDN:BFM:power-balance-P}-\ref{eq:PDN:BFM:rank-1})     \\
	& Voltage limits            &      Eqn. (\ref{eq:PDN:voltage-limits})     \\
	& Power injections          &     Eqn. (\ref{eq:PDN:DER-feasibility}-\ref{eq:PDN:power-injection-q})      \\ \hline
	\multirow{7}{*}{Water network}  & Flow rate limits          &      Eqn. (\ref{eq:MWN:pipe-limit})     \\
	& Flow continuity           &    Eqn. (\ref{eq:MWN:junction:continuity},\ref{eq:MWN:junction:minimum-head})       \\
	& Tanks                     &      Eqn. (\ref{eq:MWN:tank:head-change},\ref{eq:MWN:tank:head-bound})     \\
	& Reservoirs                &       Eqn. (\ref{eq:MWN:reservoir:head})    \\
	& Flow pumps                &    Eqn. (\ref{eq:MWN:pump-head-gain},\ref{eq:MWN:pump-power})       \\
	& Pressure valves           &       Eqn. (\ref{eq:MWN:valve-head-drop})    \\
	& Hydraulic head loss       &    Eqn. (\ref{eq:MWN:Darcy-Weisbach})       \\ \hline
	\multirow{8}{*}{Heating network} & CHPs                      &     Eqn. (\ref{eq:CHP:powergen:p}-\ref{eq:CHP:pump-power})           \\
	& Heating loads             &    Eqn. (\ref{eq:heatload:heating}-\ref{eq:heatload:min-head-drop})       \\
	& Temperature continuity    &     Eqn. (\ref{eq:DHN:heatjunction:temperature},\ref{eq:DHN:heatjunction:head})      \\
	& Hydraulic head loss       &      Eqn. (\ref{eq:DHN:Darcy-Weisbach:heatpipe:S},\ref{eq:DHN:Darcy-Weisbach:heatpipe:R})     \\
	& Flow rate limits   &     Eqn. (\ref{eq:DHN:heatpipe-limit:S},\ref{eq:DHN:heatpipe-limit:R})      \\
	& Temperature propagation   &      Eqn. (\ref{eq:thermal-propagation:S},\ref{eq:thermal-propagation:R})     \\
	& Temperature of mixed flow &     Eqn. (\ref{eq:heatnode:temperature-mix:S},\ref{eq:heatnode:temperature-mix:R})      \\
	& Flow continuity           &     Eqn. (\ref{eq:heatnode:continuity:S},\ref{eq:heatnode:continuity:R})      \\ \hline
\end{tabular}
\end{table}

\subsection{Power distribution network model}
\label{subsec:power-model}

For simplicity, we use a single-phase equivalent model for a balanced power network~\cite{farivar2013branch}, while our framework can be extended to unbalanced multiphase models~\cite{zhao2017convex}.  
The power network is modeled as a \emph{directed} graph $(\N^{\textnormal{P}}, \E^{\textnormal{P}})$ with a \emph{tree} topology. Let $V_i^t$ denote the complex voltage and let $s_i^t := p_i^t + \jay q_i^t$ denote the complex power injection at node $i \in \N^{\textnormal{P}}$ at time $t \in \T$. A node indexed by $\textnormal{E0} \in \N^{\textnormal{P}}$ is the slack bus that represents the substation and has a constant voltage $V_{E0}^t\equiv|V_{E0}|$ for all $t \in \T$.
Define $\N^{\textnormal{P},+}:= \N^{\textnormal{P}}\backslash \{\textnormal{E0}\}$.  
Let $\N^{\textnormal{P,DER}}$, $\N^{\textnormal{P,pump}}$, and $\N^{\textnormal{P,CHP}}$ denote the subsets of $\N^{\textnormal{P},+}$ that connect to controllable distributed energy resources (DERs, e.g., solar photovoltaic panels that can perform power curtailment) in the power network, pumps in the water network, and CHPs in the heating network, respectively; these subsets may overlap. 

\subsubsection{Power flow}

We adopt the branch flow model \cite{farivar2013branch}:
 \begin{eqnarray}
 && \sum_{k: ki \in \E^{\textnormal{P}}}  \left(P_{ki}^t - r_{ki}\ell_{ki}^t\right) + p_i^t  =  \sum_{j:ij \in \E^{\textnormal{P}}} P_{ij}^t 
 \label{eq:PDN:BFM:power-balance-P}
 \\
 && \sum_{k: ki \in \E^{\textnormal{P}}}  \left(Q_{ki}^t - x_{ki}\ell_{ki}^t\right) + q_i^t  = \sum_{j:ij \in \E^{\textnormal{P}}}  Q_{ij}^t, \nonumber
 \\
&& \qquad\qquad\qquad\qquad\qquad\qquad  \forall i \in \N^{\textnormal{P}}, ~ t  \in \T \label{eq:PDN:BFM:power-balance-Q}
\\
&& v_i^t - v_j^t =  2\left(r_{ij} P_{ij}^t + x_{ij} Q_{ij}^t\right) - \left|z_{ij}\right|^2 \ell_{ij}^t
\label{eq:PDN:BFM:voltage-drop} \\ 
&& v_i^t  \ell_{ij}^t = \left(P_{ij}^t\right)^2 + \left(Q_{ij}^t\right)^2, \quad  \forall ij \in \E^{\textnormal{P}}, ~t\in \T \label{eq:PDN:BFM:rank-1}
\end{eqnarray}
where, for every line $ij \in \E^{\textnormal{P}}$, let $z_{ij}:=r_{ij} + \jay x_{ij}$ denote its constant series impedance, $I_{ij}^t$ is the complex current on it, and $S_{ij}^t:=P_{ij}^t + \jay Q_{ij}^t$ is the complex power sent by node $i$ onto line $ij$ at time $t$. Define $v_i^t: = |V_i^t|^2$ and $\ell_{ij}^t := |I_{ij}^t|^2$. 

\subsubsection{Voltage limits}

Nodal voltages are constrained as:
\begin{eqnarray}
\underline v_i \leq v_i^t \leq \overline v_i, \quad\forall i \in \N^{\textnormal{P},+},~t \in \T. \label{eq:PDN:voltage-limits} 
\end{eqnarray}

\subsubsection{Power injections}

Let $ p_i^{t,\textnormal{load}}$ and $ q_i^{t,\textnormal{load}}$ be the \emph{given} active and reactive load power at node $ i \in  \N^{\textnormal{P},+}$ at time $t$. 
Without loss of generality, we assume that only one controllable DER is connected to each node $i \in \N^{\textnormal{P,DER}} \subseteq \N^{\textnormal{P},+}$, whose active and reactive power injections at time $t$ are $p_i^{t,\textnormal{r}} $ and $q_i^{t,\textnormal{r}}$, respectively. The DER operations satisfy: 
\begin{eqnarray}
(p_i^{t,\textnormal{r}}, q_i^{t,\textnormal{r}}) \in \mathcal{R}_i^t \subset \mathbb{R}^2, \quad \forall i \in \N^{\textnormal{P,DER}}, ~t \in \T \label{eq:PDN:DER-feasibility}
\end{eqnarray}
where we assume $\mathcal{R}_i^t$ to be a compact convex set. 

We assume there is one water pump, whose active power consumption at time $t$ is $p_i^{t,\textnormal{w-pump}}$, supplied by each node $i \in \N^{\textnormal{P,pump}} \subseteq \N^{\textnormal{P},+}$, e.g., node E22 in Figure \ref{fig:1}. 
We assume there is one CHP at each node $i \in \N^{\textnormal{P,CHP}} \subseteq \N^{\textnormal{P},+}$, e.g., node E21 in Figure \ref{fig:1}, whose active and reactive power generations at time $t$ are $p_{i}^{t,\textnormal{gen}}$ and $q_{i}^{t,\textnormal{gen}}$, respectively.  
A water pump is installed at every CHP, whose active power consumption at time $t$ is $p_{i}^{t,\textnormal{h-pump}}$. The models of $p_i^{t,\textnormal{w-pump}}$, $p_{i}^{t,\textnormal{gen}}$, $q_{i}^{t,\textnormal{gen}}$, and $p_{i}^{t,\textnormal{h-pump}}$ will be elaborated in Sections \ref{subsec:water-model} and \ref{subsec:heat-model}.

For convenience, we define dummy variables $p_i^{t,\textnormal{r}} = q_i^{t,\textnormal{r}} \equiv 0$ for $i \in \N^{\textnormal{P},+} \backslash \N^{\textnormal{P,DER}}$; $p_i^{t,\textnormal{gen}} = q_i^{t,\textnormal{gen}} = p_{i}^{t,\textnormal{h-pump}} \equiv 0$ for $i \in \N^{\textnormal{P},+} \backslash \N^{\textnormal{P,CHP}}$; and $p_i^{t,\textnormal{w-pump}} \equiv 0$ for $i \in \N^{\textnormal{P},+} \backslash \N^{\textnormal{P,pump}}$. Then \emph{net} active and reactive power injections at every power network node (except the slack bus) can be written as:
\begin{eqnarray}
p_i^t &=& p_i^{t,\textnormal{r}} +  p_i^{t,\textnormal{gen}} - p_i^{t,\textnormal{load}} - p_i^{t,\textnormal{w-pump}} - p_i^{t,\textnormal{h-pump}} 
\label{eq:PDN:power-injection-p}\\
q_i^t &=& q_i^{t,\textnormal{r}} +  q_i^{t,\textnormal{gen}} - q_i^{t,\textnormal{load}}, \qquad\forall i \in \N^{\textnormal{P},+}, ~t\in \T.
\label{eq:PDN:power-injection-q}
\end{eqnarray}

\subsection{Municipal water network model}
\label{subsec:water-model}

 We follow \cite{fooladivanda2017energy, zamzam2018optimal} to model the water network as a \emph{directed} graph $(\N^{\textnormal{W}}, \E^{\textnormal{W}})$, where $\N^{\textnormal{W}}$ is the set of nodes and $\E^{\textnormal{W}}$ is the set of pipes. The nodes are categorized into disjoint subsets $\N^{\textnormal{W,J}}$ of junctions, $\N^{\textnormal{W,T}}$ of tanks, and $\N^{\textnormal{W,R}}$ of reservoirs. The pipes are categorized into disjoint subsets $\E^{\textnormal{W,P}}$ of pipes each having a variable-speed pump, $\E^{\textnormal{W,V}}$ of pipes each having a pressure-reducing valve, and $\E^{\textnormal{W,F}}$ of other pipes. 
In Figure \ref{fig:1}, $\N^{\textnormal{W,J}} = \left\{\textnormal{W2}, \textnormal{W4}, \textnormal{W5}\right\}$, $\N^{\textnormal{W,T}}=\{\textnormal{W3}, \textnormal{W3}'\}$, and $\N^{\textnormal{W,R}}=\{\textnormal{W1}\}$; $\E^{\textnormal{W,P}}=\{\textnormal{W2W3}\}$ (meaning the pipe connecting W2 and W3), $\E^{\textnormal{W,V}}=\{\textnormal{W1W2}\}$, and $\E^{\textnormal{W,F}} = \left\{\textnormal{W3}'\textnormal{W4}, \textnormal{W4W5}\right\}$. Note that the tank is modeled as two nodes $\textnormal{W3}$ and $\textnormal{W3}'$ to characterize the hydraulic head difference between its inlet and outlet. This difference will be further explained immediately.

Let $q_{ij}^t$ denote the water flow rate in pipe $ij \in \E^{\textnormal{W}}$ at time $t$.\footnote{We occasionally abuse notations, e.g., $q$ for flow rates and reactive power, $x$ for reactance and OPWHF variables, $z$ for impedance and auxiliary variables, to respect conventions, hopefully without causing confusion.} Assume that water flow is unidirectional in every pipe:
\begin{eqnarray}
q_{ij}^t \geq 0, \quad \forall ij \in \E^{\textnormal{W}},~t\in \T. \label{eq:MWN:pipe-limit}
\end{eqnarray} 

\subsubsection{Junctions} 

Water flow at every junction must satisfy the continuity condition as well as the \emph{given} water demand $d_i^t$:
\begin{eqnarray}
\sum_{k: ki\in\E^{\textnormal{W}}} q_{ki}^t  = d_i^t + \sum_{j: ij \in\E^{\textnormal{W}}}q_{ij}^t, \quad \forall i\in\N^{\textnormal{W,J}},~t \in \T.
\label{eq:MWN:junction:continuity}
\end{eqnarray}
The hydraulic head $h_i^t:=h_i^{t,\textnormal{pre}} + h_i^{t,\textnormal{ele}}$ at junction $i$ at time $t$ is composed of the pressure head $h_i^{t,\textnormal{pre}}$, which is a control variable, and the elevation head $h_i^{t,\textnormal{ele}}$, which is a given constant. We therefore treat $h_i^t$ as a variable that must satisfy:
\begin{eqnarray}
h_i^t \geq \underline h_i,  \quad \forall i\in\N^{\textnormal{W,J}},~t \in \T
\label{eq:MWN:junction:minimum-head} 
\end{eqnarray} 
where constant $\underline h_i$ is the summation of the elevation head and the minimum pressure head due to engineering considerations. 
  
\subsubsection{Tanks}

Considering the fact that the hydraulic heads at the inlet and outlet of a tank are different, we split the set $\N^{\textnormal{W,T}}$ of tank nodes into two subsets $\N^{\textnormal{W,T,in}}$ and $\N^{\textnormal{W,T,out}}$ with the same cardinality. Each tank is modeled as two nodes, i.e., an inlet node $i \in \N^{\textnormal{W,T,in}}$ and a corresponding outlet node $i' = \textnormal{out}(i) \in \N^{\textnormal{W,T,out}}$. There is no pipe between nodes $i$ and $i'$. The outlet head changes over time as:    
\begin{eqnarray}
h_{i'}^t &=& h_{i'}^{t-1} + \frac{\Delta t}{S_i} \left(\sum_{k: ki \in \E^{\textnormal{W}}} q_{ki}^t - \sum_{j: i'j \in  \E^{\textnormal{W}}} q_{i'j}^t \right), \nonumber \\
&& \qquad\quad \forall i \in \N^{\textnormal{W,T,in}}, ~i' = \textnormal{out}(i),~t \in \T \label{eq:MWN:tank:head-change}
\end{eqnarray}
where constant $\Delta t$ is the duration of each time slot and constant $S_i$ is the cross-sectional area of tank $i$. The initial head $h_{i'}^{0}$ is also given as a constant. 
The outlet head is bounded from below by a constant and cannot exceed the inlet head:
\begin{eqnarray}
\underline h_{i'} \leq h_{i'}^t \leq  h_i^t, \quad \forall i \in\N^{\textnormal{W,T,in}}, ~i' = \textnormal{out}(i),~t \in \T.   \label{eq:MWN:tank:head-bound}
\end{eqnarray}

\subsubsection{Reservoirs} 

We model reservoirs as infinite sources of water with constant hydraulic heads: 
\begin{eqnarray} 
h_i^t \equiv \underline h_i,\quad \forall i \in  \N^{\textnormal{W,R}},~t \in \T. 
\label{eq:MWN:reservoir:head}
\end{eqnarray}

\subsubsection{Variable-speed pumps} 

Pumps consume electric power to elevate hydraulic head along the pipes:
\begin{eqnarray}
&&0 \leq h_j^t - h_i^t \leq -A_{ij} \left(q_{ij}^t\right)^2 + B_{ij} q_{ij}^t  + C_{ij}  
\label{eq:MWN:pump-head-gain} \\
&&p_{\textnormal{wp}(ij)}^{t,\textnormal{w-pump}} =  \frac{\rho g}{\eta_{ij}} (h_j^t - h_i^t) q_{ij}^t, ~ \forall ij \in \E^{\textnormal{W,P}},  t \in \T  \label{eq:MWN:pump-power}
\end{eqnarray}
where $A_{ij}$, $B_{ij}$, $C_{ij}$ are \emph{positive} constants; $\textnormal{wp}(ij) \in \N^{\textnormal{P,pump}}$ maps pipe $ij$ (e.g., W2W3) to the power network node (e.g., E22) that supplies electricity to the pump on pipe $ij$; and $p_{\textnormal{wp}(ij)}^{t,\textnormal{w-pump}}$ is the power consumption of that pump at time $t$, which depends on constant pump efficiency $\eta_{ij}$, water density $\rho$, and gravity coefficient $g$. The power and water networks are thus coupled through \eqref{eq:PDN:power-injection-p}, \eqref{eq:MWN:pump-power}.

\subsubsection{Pressure-reducing valves}

The hydraulic head drops by a controllable amount along a pipe with a valve:
\begin{eqnarray}
h_j^t \leq h_i^t, \quad \forall ij \in \E^{\textnormal{W,V}}, ~ t \in \T.  \label{eq:MWN:valve-head-drop}
\end{eqnarray}

\subsubsection{Other pipes}

Hydraulic heads and flow rates along other pipes satisfy the Darcy-Weisbach equation:
\begin{eqnarray}
h_i^t - h_j^t = F_{ij}\left(q_{ij}^t\right)^2, \quad \forall ij \in \E^{\textnormal{W,F}}, ~t \in \T  \label{eq:MWN:Darcy-Weisbach}
\end{eqnarray}
where $F_{ij}>0$ is the constant friction factor of pipe $ij$.

\subsection{District heating network model}
\label{subsec:heat-model}

We set up a model by consolidating those from \cite{lahdelma2003CHP, sandou2005DHopt, chen2015DHopt, li2016DHoptLinear, li2016DHoptNonlinear}.
The heating network is represented by a \emph{directed} graph $(\N^{\textnormal{H}}, \E^{\textnormal{H}})$, where $\N^{\textnormal{H}}$ is the set of nodes and $\E^{\textnormal{H}}$ is the set of pipes. 
Each node has a supply side and a return side. For instance, in Figure \ref{fig:1}, the supply side and return side of node H1 are denoted by S1 and R1, respectively. The node set $\N^{\textnormal{H}}$ is categorized into disjoint subsets $\N^{\textnormal{H,G}}$ of CHPs, $\N^{\textnormal{H,L}}$ of heating loads, and $\N^{\textnormal{H,J}}$ of junctions.
The pipe set $\E^{\textnormal{H}}$ is categorized into disjoint subsets $\E^{\textnormal{H,S}}$ of supply pipes and $\E^{\textnormal{H,R}}$ of return pipes. A pump is installed in every CHP. Water is heated at CHPs and pumped to heating loads through supply pipes. After transferring heat to loads, water flows back to CHPs through return pipes and enters the next cycle.

\subsubsection{CHPs}

A convex polyhedron model is utilized to characterize CHP electricity and heat generation. The CHP efficiency is not explicitly modeled but is reflected in its operational cost, i.e., the second term $C_{i,k} \alpha^t_{i,k}$ in the objective function ${C_{\textnormal{total}}}$ at the beginning of Section \ref{sec:problem}, which is a convex combination of the costs under extreme operational conditions of the CHP. This model and its parameters are adopted from \cite{lahdelma2003CHP}:
\begin{eqnarray}
p_{i}^{t,\textnormal{gen}} &=& \sum_{k \in \K_i } \alpha^t_{i,k} p_{i,k}, \quad \forall i \in \N^{\textnormal{H,G}}, ~t \in \T \label{eq:CHP:powergen:p} \\
q_{i}^{t,\textnormal{gen}} &=& \sum_{k \in \K_i}  \alpha^t_{i,k} q_{i,k}, \quad \forall i \in \N^{\textnormal{H,G}}, ~t \in \T \label{eq:CHP:powergen:q} \\
H_{i}^{t,\textnormal{gen}} &=& \sum_{k \in \K_i} \alpha^t_{i,k} H_{i,k}, \quad \forall i \in \N^{\textnormal{H,G}}, ~t \in \T \label{eq:CHP:heatgen}
\end{eqnarray}
where $\K_i:=\left\{1,2,\dots,|\K_i| \right\}$ is the index set for constant extreme points $(p_{i,k}, q_{i,k}, H_{i,k})$; active and reactive power generations $(p_{i}^{t,\textnormal{gen}}, q_{i}^{t,\textnormal{gen}})$ couple the power and heating networks through \eqref{eq:PDN:power-injection-p}, \eqref{eq:PDN:power-injection-q}, \eqref{eq:CHP:powergen:p}, \eqref{eq:CHP:powergen:q}; and $H_{i}^{t,\textnormal{gen}}$ is the thermal power generated by CHP $i$ at time $t$. Decision variables $\alpha^t_{i,k}$ satisfy:
\begin{eqnarray}
&&\sum_{k\in \K_i} \alpha^t_{i,k} = 1, \quad \forall i \in \N^{\textnormal{H,G}},~ t \in \T \label{eq:CHP:convex-1} \\
&&0 \leq \alpha^t_{i,k}\leq 1, \quad \forall i \in \N^{\textnormal{H,G}},~t \in \T, ~k \in \K_i. \IEEEeqnarraynumspace \label{eq:CHP:convex-2}
\end{eqnarray} 
The heat generated by CHPs is used to heat the water:
\begin{eqnarray}
H_{i}^{t,\textnormal{gen}} = c \cdot q_i^{t,\textnormal{RS}} (\tau_i^{t,\textnormal{S}} - \tau_i^{t,\textnormal{R}}), \quad \forall i \in \N^{\textnormal{H,G}}, ~ t \in \T \label{eq:CHP:heating} 
\end{eqnarray}
where constant $c$ is the product of specific heat capacity and density of water; $\tau_i^{t,\textnormal{S}}$ and $\tau_i^{t,\textnormal{R}}$ are supply and return temperatures, respectively; $q_i^{t,\textnormal{RS}}$ is the water flow rate from the return side to the supply side of CHP $i$ at time $t$. The supply temperature is bounded as:
\begin{eqnarray}
\underline \tau_i^{\textnormal{S}} \leq \tau_i^{t,\textnormal{S}} \leq \overline \tau_i^{\textnormal{S}},\quad \forall i \in \N^{\textnormal{H,G}}, ~ t \in \T. \label{eq:CHP:supplyTbound}
\end{eqnarray}
We assume that water flow is unidirectional in every CHP:
\begin{eqnarray}
q_i^{t,\textnormal{RS}} \geq 0, \quad \forall i \in \N^{\textnormal{H,G}},~t \in \T. \label{eq:DHN:CHP:flow-limit}
\end{eqnarray} 
The pump in every CHP consumes electricity to elevate hydraulic head from the return side to the supply side:
\begin{eqnarray}
&&0 \leq h_i^{t,\textnormal{S}} - h_i^{t,\textnormal{R}} \leq -A_{i} \left(q_i^{t,\textnormal{RS}}\right)^2 + B_{i} q_i^{t,\textnormal{RS}}  + C_i  
\label{eq:CHP:pump-head-gain} \\
&&p_{\textnormal{hp}(i)}^{t,\textnormal{h-pump}} =  \frac{\rho g}{\eta_{i}} (h_i^{t,\textnormal{S}} - h_i^{t,\textnormal{R}}) q_i^{t,\textnormal{RS}}, ~ \forall i \in \N^{\textnormal{H,G}},  t \in \T \IEEEeqnarraynumspace \label{eq:CHP:pump-power}
\end{eqnarray}
where $A_i$, $B_i$, $C_i$ are \emph{positive} constants; $\textnormal{hp}(i) \in \N^{\textnormal{P,CHP}}$ maps CHP node $i$ (e.g., H1) to the power network node (e.g., E21) that supplies electricity to the pump in that CHP; $h_i^{t,\textnormal{S}}$ and $h_i^{t,\textnormal{R}}$ are hydraulic heads at the supply and return sides, respectively. Active power consumption $p_{\textnormal{hp}(i)}^{t,\textnormal{h-pump}}$ of the pump couples the power and heating networks through \eqref{eq:PDN:power-injection-p}, \eqref{eq:CHP:pump-power}.

\subsubsection{Heating loads}
 
Water temperature drops at loads:
\begin{eqnarray}
H_{i}^{t,\textnormal{load}} = c \cdot  q_i^{t,\textnormal{RS}} (\tau_i^{t,\textnormal{R}} - \tau_i^{t,\textnormal{S}}), \quad \forall i \in \N^{\textnormal{H,L}}, ~ t \in \T \label{eq:heatload:heating} 
\end{eqnarray}
so that a \emph{given} thermal power $H_{i}^{t,\textnormal{load}}\geq 0$ is delivered to load $i$ at time $t$. Water temperatures at the supply side and return side of a load node $i \in\N^{\textnormal{H},\textnormal{L}}$ at time $t$ are denoted by $\tau_i^{t,\textnormal{S}}$ and $\tau_i^{t,\textnormal{R}}$, respectively. 
The return temperature is bounded:
\begin{eqnarray}
\underline \tau_i^{\textnormal{R}} \leq \tau_i^{t,\textnormal{R}} \leq \overline \tau_i^{\textnormal{R}},\quad \forall i \in \N^{\textnormal{H,L}}, ~ t \in \T. \label{eq:heatload:returnTbound}
\end{eqnarray}
Water flow in a heating load is unidirectional from the supply side to the return side, which, represented in $q_i^{t,\textnormal{RS}}$, satisfies: 
\begin{eqnarray}
q_i^{t,\textnormal{RS}} \leq 0, \quad \forall i \in \N^{\textnormal{H,L}},~t \in \T.  \label{eq:heatload:flow-limit}
\end{eqnarray} 
Moreover, the hydraulic head at a heating load is required to drop by a minimum amount to sustain the flow:
\begin{eqnarray}
h_i^{t,\textnormal{S}} - h_i^{t,\textnormal{R}} \geq \underline h_i^{\textnormal{SR}}, \quad \forall i \in \N^{\textnormal{H,L}}, ~ t \in \T.
\label{eq:heatload:min-head-drop}
\end{eqnarray}

\subsubsection{Junctions} 
 
A junction neither generates nor consumes heat, so its supply and return temperatures are the same:
\begin{eqnarray}
\tau_i^{t,\textnormal{S}} =  \tau_i^{t,\textnormal{R}}, \quad \forall i \in \N^{\textnormal{H,J}}, ~ t \in \T. 
\label{eq:DHN:heatjunction:temperature} 
\end{eqnarray}
The change of hydraulic head at a junction is also negligible:
\begin{eqnarray}
h_i^{t,\textnormal{S}} =  h_i^{t,\textnormal{R}}, \quad \forall i \in \N^{\textnormal{H,J}}, ~ t \in \T.
\label{eq:DHN:heatjunction:head}
\end{eqnarray}

\subsubsection{Pipes} 

For every supply pipe $ij \in \E^{\textnormal{H,S}}$ with water flow direction $i\rightarrow j$, there is also a return pipe $ji \in \E^{\textnormal{H,R}}$ with water flow direction $j \rightarrow i$. We assume there is no pump on any pipe, and hence the hydraulic heads and flow rates through the pipes are related by Darcy-Weisbach equations:
\begin{eqnarray}
h_i^{t,\textnormal{S}} - h_j^{t,\textnormal{S}} &=& F^{\textnormal{S}}_{ij} \left(q_{ij}^{t,\textnormal{S}} \right)^2, ~ \forall  ij \in \E^{\textnormal{H,S}}, t \in \T 
\label{eq:DHN:Darcy-Weisbach:heatpipe:S}\\
h_j^{t,\textnormal{R}} - h_i^{t,\textnormal{R}} &=& F^{\textnormal{R}}_{ji} \left(q_{ji}^{t,\textnormal{R}} \right)^2, ~ \forall ji \in \E^{\textnormal{H,R}}, t \in \T
\label{eq:DHN:Darcy-Weisbach:heatpipe:R}
\end{eqnarray}  
where $h_i^{t,\textnormal{S}}$ and $h_j^{t,\textnormal{S}}$ ($h_j^{t,\textnormal{R}}$ and $h_i^{t,\textnormal{R}}$) are hydraulic heads at the sending and receiving nodes of supply (return) pipe $ij$ ($ji$), respectively; $q_{ij}^{t,\textnormal{S}}$ ($q_{ji}^{t,\textnormal{R}}$) is the flow rate in supply (return) pipe $ij$ ($ji$) at time $t$; and $F^{\textnormal{S}}_{ij}$ ($F^{\textnormal{R}}_{ji}$) is the constant friction factor of supply (return) pipe $ij$ ($ji$). 
Pipe flow is unidirectional:
\begin{eqnarray}
&& q_{ij}^{t,\textnormal{S}} \geq 0,\quad \forall ij \in \E^{\textnormal{H,S}}, ~t \in \T 
\label{eq:DHN:heatpipe-limit:S}\\
&& q_{ji}^{t,\textnormal{R}} \geq 0,\quad \forall ji \in \E^{\textnormal{H,R}}, ~t \in \T. 
\label{eq:DHN:heatpipe-limit:R}
\end{eqnarray} 
Assuming the timescale of thermal dynamics in the heating network is much faster than OPWHF, we consider the steady state of the temperature propagation model in \cite{sandou2005DHopt}:
\begin{eqnarray}
&&\tau_{ij}^{t,\textnormal{S,out}}  =   \left( \tau_{i}^{t,\textnormal{S}}\! -\! \tau_0^t \right) e^{-\frac{\xi_{ij}^\textnormal{S}}{q_{ij}^{t,\textnormal{S}}}} \!+\! \tau_0^t, ~ \forall ij \in \E^{\textnormal{H,S}}, t\in\T \IEEEeqnarraynumspace \label{eq:thermal-propagation:S}\\
&&\tau_{ji}^{t,\textnormal{R,out}}  =   \left( \tau_{j}^{t,\textnormal{R}} \!-\!\tau_0^t \right) e^{-\frac{\xi_{ji}^\textnormal{R}}{q_{ji}^{t,\textnormal{R}}}} \!+\! \tau_0^t, ~ \forall ji \in \E^{\textnormal{H,R}}, t\in\T \IEEEeqnarraynumspace \label{eq:thermal-propagation:R}
\end{eqnarray}  
where $\tau_0^t$ is the ambient temperature at time $t$ and $\xi_{ij}^\textnormal{S}$ ($\xi_{ji}^\textnormal{R}$) is a constant parameter determined by the length and diameter of supply (return) pipe $ij$ ($ji$) as well as the constant heat transfer coefficient, density, and specific heat capacity of water \cite{sandou2005DHopt}. Moreover, $\tau_{ij}^{t,\textnormal{S,out}}$ ($\tau_{ji}^{t,\textnormal{R,out}}$) is the temperature of water flowing out of supply (return) pipe $ij$ ($ji$). Water flowing out of different pipes into the same node is mixed, and its temperature as a result of energy conservation is:
\begin{eqnarray}
\sum_{k:ki\in\E^{\textnormal{H,S}}} \left(q_{ki}^{t,\textnormal{S}} \tau_{ki}^{t,\textnormal{S,out}}\right) &=& \left(\sum_{k:ki\in\E^{\textnormal{H,S}}} q_{ki}^{t,\textnormal{S}} \right) \tau_{i}^{t,\textnormal{S}}
\label{eq:heatnode:temperature-mix:S}\\
\sum_{j:ji\in\E^{\textnormal{H,R}}} \left(q_{ji}^{t,\textnormal{R}} \tau_{ji}^{t,\textnormal{R,out}}\right)  &=& \left(\sum_{j:ji\in\E^{\textnormal{H,R}}} q_{ji}^{t,\textnormal{R}} \right) \tau_{i}^{t,\textnormal{R}},
\nonumber \\
&&\qquad \forall i \in \N^{\textnormal{H}}, ~t \in \T.   
\label{eq:heatnode:temperature-mix:R}
\end{eqnarray}
Last, flow at every node must satisfy the continuity condition:
\begin{eqnarray}
\hspace{-1em} \sum_{k:ki\in\E^{\textnormal{H,S}}} q_{ki}^{t,\textnormal{S}} + q_i^{t,\textnormal{RS}} = \sum_{j:ij \in \E^{\textnormal{H,S}}} q_{ij}^{t,\textnormal{S}}, && \hspace{-1.4em}  \forall i \in \N^{\textnormal{H}}, t \in \T
\label{eq:heatnode:continuity:S}\\
\hspace{-1em} \sum_{j:ji\in\E^{\textnormal{H,R}}} q_{ji}^{t,\textnormal{R}} -  q_i^{t,\textnormal{RS}} = \sum_{k:ik \in\E^{\textnormal{H,R}}} q_{ik}^{t,\textnormal{R}} ,
&& \hspace{-1.4em}  \forall i \in \N^{\textnormal{H}}, t \in \T.   \label{eq:heatnode:continuity:R}
\end{eqnarray}

\section{Optimal Power-Water-Heat Flow Problem}\label{sec:problem}

We consider the following operational cost of the interconnected power-water-heating network:
\begin{eqnarray}
 &&C_{\textnormal{total}} (p^{\textnormal{r}}, \alpha, p_0) :=\sum_{t \in \T} \bigg(\sum_{i \in \N^{\textnormal{P,DER}}} C_i^{t, \textnormal{r}} (p_i^{t,\textnormal{r}})   \nonumber
\\
& &\qquad + \sum_{i \in \N^{\textnormal{H,G}}} \sum_{k \in \K_i} C_{i,k} \alpha^t_{i,k} + \lambda_{\textnormal{power}}^t  p_{\textnormal{E0}}^t + \lambda_{\textnormal{water}}^t  q_{\textnormal{W0}}^t\bigg) \nonumber
\end{eqnarray}
where $p^{\textnormal{r}}$, $\alpha$, $p_0$ denote the vectors collecting $p_i^{t,\textnormal{r}}$, $\alpha^t_{i,k}$, $p_0^t$, respectively, over appropriate indices $i$, $k$, and $t$. Specifically:
\begin{itemize}
\item $C_i^{t, \textnormal{r}}$ is the operational cost of DER $i$ at time $t$, which is determined by the DER active power injection $p_i^{t,\textnormal{r}}$. We assume $C_i^{t, \textnormal{r}}$ are convex functions for $i \in \N^{\textnormal{P,DER}}$, $t \in \T$. 

\item The operational cost of CHP $i$ at time $t$ is modeled as the convex combination $\sum_{k \in \K_i} C_{i,k} \alpha^t_{i,k}$ of the constant costs $C_{i,k}$ at the extreme operating points \cite{lahdelma2003CHP,li2016DHoptLinear}.

\item The total electricity consumption of the power, water, and heating networks, including that for the electric loads, power losses, and water pumps after utilizing the local power supply from CHPs, is supplied by the active power flow $p_{\textnormal{E0}}^t$ through the power substation $\textnormal{E0}$ at time $t$, so that given electricity price $\lambda_{\textnormal{power}}^t$, the total electricity payment of the interconnected system is $\lambda_{\textnormal{power}}^t  p_{\textnormal{E0}}^t$ at time $t$. Let $q_{\textnormal{W0}}^t$ denote the total water supply from the reservoir into the water network, so that given water price $\lambda_{\textnormal{water}}^t$, the payment for water is $  \lambda_{\textnormal{water}}^t  q_{\textnormal{W0}}^t$ at time $t$. 
\end{itemize}

We group all the variables in Section \ref{sec:model} as follows:
\begin{itemize}
\item In power network $x_{\textnormal{power}}:=(p, q, P, Q, \ell, v, p^{\textnormal{r}}, q^{\textnormal{r}})$.

\item In water network $x_{\textnormal{water}}:=(q,h)$.

\item In heating network
\begin{eqnarray}
x_{\textnormal{heat}}:=(\alpha, \! H^{\textnormal{gen}}, \! q^{\textnormal{RS}}, \!q^{\textnormal{S}},\! q^{\textnormal{R}},\! h^{\textnormal{S}},\! h^{\textnormal{R}}, \!\tau^{\textnormal{S}},\! \tau^{\textnormal{R}},\!  \tau^{\textnormal{S,out}}, \!  \tau^{\textnormal{R,out}}).
 \nonumber
\end{eqnarray}

\item Coupling variables $x_{\textnormal{couple}}:=(p^{\textnormal{gen}}, q^{\textnormal{gen}}, p^{\textnormal{w-pump}}, p^{\textnormal{h-pump}})$.
\end{itemize}

With $x:=(x_{\textnormal{power}}, x_{\textnormal{water}}, x_{\textnormal{heat}}, x_{\textnormal{couple}})$, we formulate an optimal power-water-heat flow (OPWHF) problem as:
\begin{eqnarray}
\textbf{OPWHF:}\quad \min_x &~& C_{\textnormal{total}}(x) 
\nonumber
 \\
\textnormal{subject to} &~& \eqref{eq:PDN:BFM:power-balance-P}\text{--}\eqref{eq:heatnode:continuity:R}.  \nonumber 
\end{eqnarray}

\section{Solution Method}\label{sec:solution}

We present a heuristic to solve the nonconvex OPWHF problem that involves complex couplings between electric, hydraulic, and thermal models above. The idea behind the proposed heuristic is sketched as follows.       

First, OPWHF is approximated with convex relaxation and convex-concave procedure. The convex-concave procedure is iterative, with a new convex problem derived every iteration based on the optimal solution of the last iteration.     

Second, a multi-energy integrated system may be operated in different sectors, each having its own private information. This motivates us to propose a distributed/decomposed solution process for the OPWHF problem. The distributed process can conveniently deal with the interdependencies and interactions between different sectors, while reducing the required amount of information they have to disclose to a central operator. Inspired by \cite{geoffrion1972generalized}, we decompose the convex approximation of OPWHF into three sectors, namely power, water, and heat. Iteratively, each sector solves a smaller subproblem while the three sectors exchange information to reach consensus on coupling variables. 

Third, in the traditional setting, decomposed subproblems are solved iteratively in every iteration of convex-concave procedure, i.e., convex-concave procedure is the outer loop and decomposition is the inner loop. To reduce information exchange burdens between operators, we swap decomposition to the outer loop and convex-concave procedure to the inner loop. Specifically, convex-concave procedure is carried out iteratively within each subproblem until convergence, at which time information is exchanged between subproblems. 
Therefore, unlike \cite{geoffrion1972generalized} where decomposition was performed directly on a convex problem, our modified decomposition is on a nonconvex problem (since it is applied in the outer loop before the problem is convexified), with separate iterative convex approximation to subproblems. 

A possible risk of the proposed distributed process may be its slower convergence than a centralized method. We show in Section \ref{subsec:simulation:performance} that the proposed algorithm can converge within reasonable time for OPWHF implementation.
The rigorous proof of its convergence and optimality is a challenging task left for future work.
We now elaborate our method.

\subsection{Convex approximation}

The following convex approximations are applied to various nonconvex constraints in OPWHF.  

\subsubsection{Power flow}

The nonconvex quadratic equation \eqref{eq:PDN:BFM:rank-1} can be relaxed to a convex second-order cone (SOC)  \cite{farivar2013branch}:
\begin{eqnarray}
 v_i^t  \ell_{ij}^t \geq \left(P_{ij}^t\right)^2 + \left(Q_{ij}^t\right)^2, \quad  \forall ij \in \E^{\textnormal{P}}, ~t\in \T.    \label{eq:PDN:BFM:soc}
\end{eqnarray} 

\subsubsection{Darcy-Weisbach equations}

The nonconvex quadratic equations \eqref{eq:MWN:Darcy-Weisbach}, \eqref{eq:DHN:Darcy-Weisbach:heatpipe:S}, \eqref{eq:DHN:Darcy-Weisbach:heatpipe:R} can be relaxed to SOCs \cite{fooladivanda2017energy}:
\begin{eqnarray}
&& h_i^t - h_j^t= F_{ij} W_{ij}^t, \qquad \left(q_{ij}^t\right)^2 \leq W_{ij}^t 
\nonumber \\
&& \left(W_{ij}^t\right)^2 \leq \gamma_{ij}^t  q_{ij}^t,~\forall ij \in \E^{\textnormal{W,F}} \!\cup\! \E^{\textnormal{H,S}} \!\cup\! \E^{\textnormal{H,R}}, t \in \T \IEEEeqnarraynumspace  \label{eq:Darcy-Weisbach:soc}
\end{eqnarray} 
where $W_{ij}^t$ and $\gamma_{ij}^t$ are auxiliary variables. We skip superscripts $\textnormal{S}$ and $\textnormal{R}$ in \eqref{eq:Darcy-Weisbach:soc} for simplicity of notation.

\subsubsection{Bilinear terms} 

Constraints \eqref{eq:MWN:pump-power}, \eqref{eq:CHP:heating}, \eqref{eq:CHP:pump-power}, \eqref{eq:heatload:heating}, \eqref{eq:heatnode:temperature-mix:S}, \eqref{eq:heatnode:temperature-mix:R} are nonconvex because of the bilinear terms, i.e., the products of $(h, q)$ or $(\tau, q)$. Take \eqref{eq:MWN:pump-power} for example, which can be equivalently written as:
\begin{eqnarray}
&&p_{\textnormal{wp}(ij)}^{t,\textnormal{w-pump}} =  \frac{\rho g}{\eta_{ij}} \left(z_{ij,j}^t - z_{ij,i}^t\right), ~ \forall ij \in \E^{\textnormal{W,P}},  t \in \T  \IEEEeqnarraynumspace \label{eq:MWN:pump-power:linear} 
\\
&& z_{ij,j}^t = h_{j}^t q_{ij}^t, \quad z_{ij,i}^t = h_{i}^t q_{ij}^t,~ \forall ij \in \E^{\textnormal{W,P}}, t \in \T  \label{eq:MWN:pump-power:bilinear}
\end{eqnarray} 
where $z_{ij,j}^t$ and $z_{ij,i}^t$ are auxiliary variables making \eqref{eq:MWN:pump-power:linear} a linear constraint. The nonconvex constraint $z_{ij,j}^t = h_{j}^t q_{ij}^t$ in \eqref{eq:MWN:pump-power:bilinear} can be further transformed to:
\begin{eqnarray}
\frac{1}{2} \left(h_{j}^t + q_{ij}^t\right)^2 - \frac{1}{2}\left((h_{j}^t)^2 + (q_{ij}^t)^2\right) &\leq& z_{ij,j}^t 
\label{eq:MWN:pump-power:convex-concave-1}
\\
\frac{1}{2}\left((h_{j}^t)^2 + (q_{ij}^t)^2\right)  -  \frac{1}{2} \left(h_{j}^t + q_{ij}^t\right)^2  &\leq& - z_{ij,j}^t. 
\label{eq:MWN:pump-power:convex-concave-2}
\end{eqnarray}  
For each of \eqref{eq:MWN:pump-power:convex-concave-1} and \eqref{eq:MWN:pump-power:convex-concave-2}, the first term on the left-hand-side is a convex function and the second term (including the minus sign) is a concave function of $(h_{j}^t, q_{ij}^t)$. Given a constant reference point $(h_{j}^{t,o}, q_{ij}^{t,o})$, we perform the convex-concave procedure \cite{lipp2016variations} to linearize the concave components in \eqref{eq:MWN:pump-power:convex-concave-1} and \eqref{eq:MWN:pump-power:convex-concave-2} to obtain their convex approximates: 
\begin{eqnarray}
&&\frac{1}{2} \left(h_{j}^t + q_{ij}^t\right)^2 - \frac{1}{2}\left(( h_{j}^{t,o})^2 + (q_{ij}^{t,o})^2\right) \nonumber
\\
 &&\quad -   h_{j}^{t,o} (h_j^t -   h_j^{t,o}) -   q_{ij}^{t,o} (q_{ij}^t-   q_{ij}^{t,o}) \leq z_{ij,j}^t 
\label{eq:MWN:pump-power:convex-1}
\\
&& \frac{1}{2}\left((h_{j}^t)^2 + (q_{ij}^t)^2\right)  -  \frac{1}{2} \left(h_{j}^{t,o} +   q_{ij}^{t,o}\right)^2 \nonumber 
\\
  && \quad - (h_{j}^{t,o} +   q_{ij}^{t,o}) (h_{j}^t -   h_{j}^{t,o} +  q_{ij}^t - q_{ij}^{t,o}) \leq - z_{ij,j}^t. \IEEEeqnarraynumspace
\label{eq:MWN:pump-power:convex-2}
\end{eqnarray}  
The same approximation applies to $z_{ij,i}^t = h_{i}^t q_{ij}^t$. The same procedure as \eqref{eq:MWN:pump-power:linear}--\eqref{eq:MWN:pump-power:convex-2} can be implemented to obtain convex approximates of \eqref{eq:CHP:heating}, \eqref{eq:CHP:pump-power}, \eqref{eq:heatload:heating}, \eqref{eq:heatnode:temperature-mix:S},  \eqref{eq:heatnode:temperature-mix:R}. 

\subsubsection{Temperature propagation equations}

In \eqref{eq:thermal-propagation:S}, \eqref{eq:thermal-propagation:R}, variable flow rates $q$ appear in the denominators of exponents. We consider these constraints in simplified notations:
\begin{eqnarray}
\tau^{\textnormal{out}}   =   \left( \tau -\tau_0 \right) e^{-\frac{\xi}{q}} + \tau_0 \nonumber
\end{eqnarray}
which can be equivalently written as:
\begin{eqnarray}
  \tau y  = \tau^{\textnormal{out}}   - \tau_0 + \tau_0 y, &~&  q z = -\xi 
\label{eq:thermal-propagation:bilinear}
\\
\ln(y) &\geq& z
\label{eq:thermal-propagation:log:convex}
\\ 
\ln(y) &\leq& z
\label{eq:thermal-propagation:log:concave} 
\end{eqnarray}
where $y$ and $z$ are auxiliary variables to substitute $e^{-\frac{\xi}{q}}$ and $-\xi/q$, respectively. Bilinear constraints \eqref{eq:thermal-propagation:bilinear} can be treated with the same convex-concave procedure as \eqref{eq:MWN:pump-power:bilinear}--\eqref{eq:MWN:pump-power:convex-2}. Constraint \eqref{eq:thermal-propagation:log:convex} is convex while \eqref{eq:thermal-propagation:log:concave} is not. Around a given point $y^o \in (0,1)$, we linearize $\ln(y)$ to approximate \eqref{eq:thermal-propagation:log:concave} as:
\begin{eqnarray}
\ln(y^o) + \frac{1}{y^o} (y - y^o) \leq z. \label{eq:thermal-propagation:log:linearize}
\end{eqnarray}

There may be effective alternative methods to deal with the nonlinear, nonconvex  constraints. One example is to piecewisely linearize the nonlinear functions using integer variables to obtain a mixed integer SOC program. We choose the convex-concave procedure due to its following advantages: 
\begin{enumerate}
\item It does not require additional integer variables.
\item It only linearizes the concave component of a nonconvex constraint, while retaining all the higher-order information in its convex component. 
\item It provides a convex inner approximation to a nonconvex feasible set and hence guarantees feasibility of the obtained solution.
\end{enumerate}

\subsection{Problem decomposition}\label{subsec:decomposed_solution}

\begin{algorithm}[!t]
	{\bf Initialize} outer iteration index $m=0$. Find $\tilde x_{\textnormal{water}}^{*,0}$ and $\tilde x_{\textnormal{heat}}^{*,0}$ that are feasible for nonconvex OWF and OHF given $x_{\textnormal{couple}}^{*,0}$.
	
	{\bf Repeat}	
	\begin{enumerate}[label={},  leftmargin = 1em]
		\item \textbf{- Step 0.} Reset inner iteration indices $n_w=0$, $n_h=0$. Reset reference points $\tilde x_{\textnormal{water}}^{o,0} = \tilde x_{\textnormal{water}}^{*,m}$ and $\tilde x_{\textnormal{heat}}^{o,0} = \tilde x_{\textnormal{heat}}^{*,m}$.  
		\item \textbf{- Step 1. Repeat}
		\begin{enumerate}[leftmargin = 2em]
			\item Solve $\textnormal{OWF-C}(x_\textnormal{couple}^{*,m}, \tilde x_\textnormal{water}^{o,n_w})$; denote the optimal solution as $\tilde x_\textnormal{water}^{o,n_w+1}$; make $n_w \leftarrow n_w+1$.
			\item Solve $\textnormal{OHF-C}(x_\textnormal{couple}^{*,m}, \tilde x_\textnormal{heat}^{o,n_h})$; denote the optimal solution as $\tilde x_\textnormal{heat}^{o,n_h+1}$; make $n_h \leftarrow n_h+1$.
		\end{enumerate}
		\textbf{Until} inner convergence criterion is met for Step 1. 
		\item \textbf{- Step 2.} Make $m \leftarrow m+1$. Update $\tilde x_{\textnormal{water}}^{*,m} = \tilde x_{\textnormal{water}}^{o,n_w}$ and $\tilde x_{\textnormal{heat}}^{*,m} = \tilde x_{\textnormal{heat}}^{o,n_h}$. Record optimal dual variables from the last iteration of Step 1 as $\mu^{*,m} := (\mu_\textnormal{water}^{*,m}, \mu_\textnormal{heat}^{*,m})$. Also record:
		\begin{eqnarray}\nonumber
		G^{*,m}:=  \left(G_\textnormal{water}(\tilde x_\textnormal{water}^{o,n_w-1}; \tilde x_\textnormal{water}^{*,m}), G_\textnormal{heat}(\tilde x_\textnormal{heat}^{o,n_h-1}; \tilde x_\textnormal{heat}^{*,m})\right).
		\end{eqnarray}
		\item \textbf{- Step 3.} Solve $\textnormal{OPF-C}(\mu^{*,m} , G^{*,m})$ to get $(x_\textnormal{power}^{*,m}, x_\textnormal{couple}^{*,m})$.
	\end{enumerate}
	
	{\bf Until} outer convergence criterion is met.
	
	{\bf Output} $x^*:=(x_{\textnormal{power}}^{*,m}, x_{\textnormal{water}}^{*,m}, x_{\textnormal{heat}}^{*,m}, x_{\textnormal{couple}}^{*,m})$.
	\caption{Decomposed algorithm to solve OPWHF.}
	\label{alg:decomposed}
\end{algorithm}

\begin{figure}
	\includegraphics[width=0.85\columnwidth]{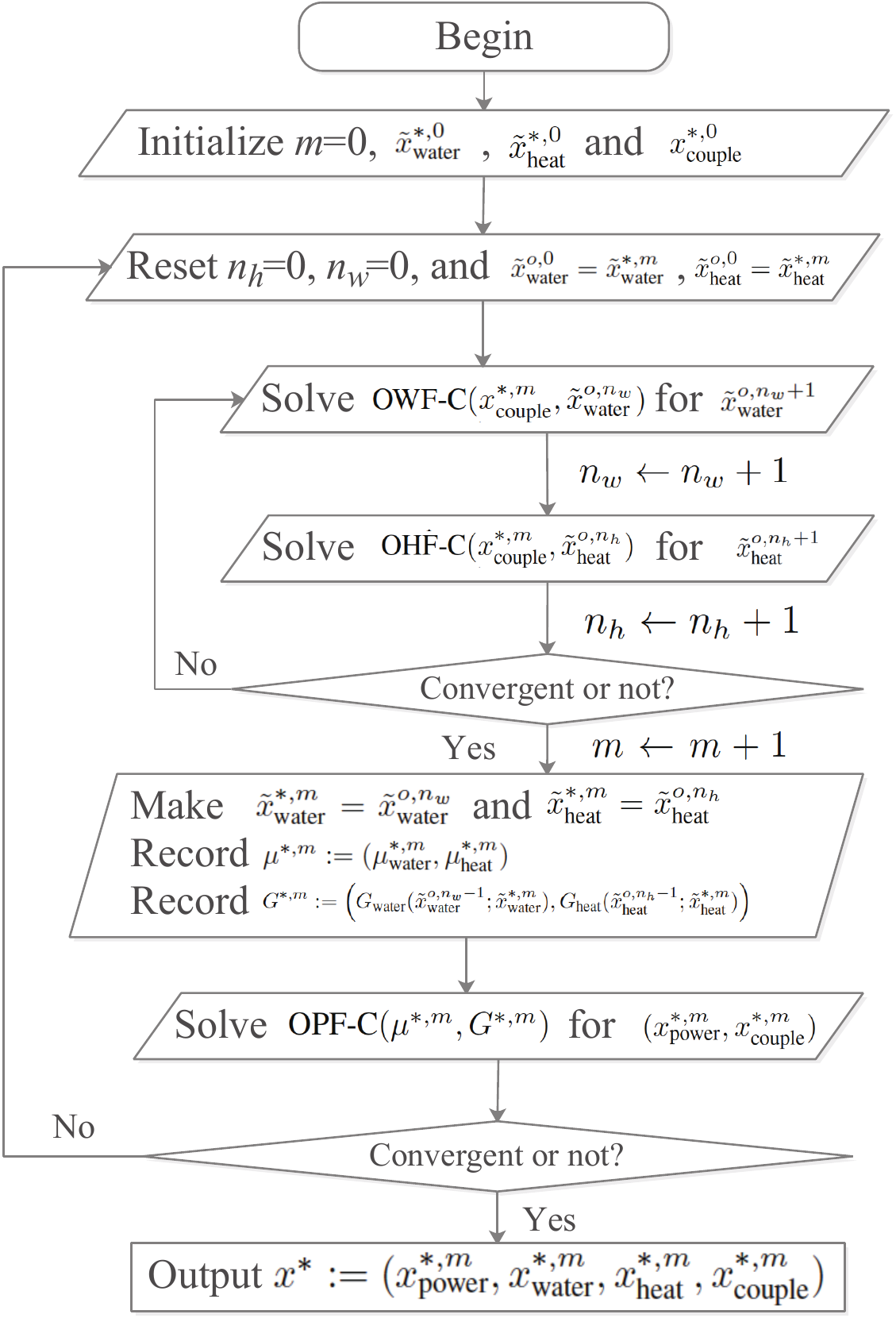}
	\centering
	\caption{Flowchart of Algorithm \ref{alg:decomposed}.}
	\label{fig:flowchart}
\end{figure}

We decompose OPWHF and apply convex approximations above to obtain three \emph{convex} subproblems: convex optimal power flow (OPF-C), convex optimal water flow (OWF-C), and convex optimal heat flow (OHF-C). 
\begin{eqnarray}
\textbf{OPF-C}: && \min_{x_\textnormal{power}, x_\textnormal{couple}}  \sum_{t \in \T} \left( {\sum_{i \in \N^{\textnormal{P,DER}}} C_i^{t, \textnormal{r}} (p_i^{t,\textnormal{r}})+\lambda_{\textnormal{power}}^t}{p_{\textnormal{E0}}^t}\right)  \nonumber \\
&& \qquad  \qquad +\left(\mu^*\right)^T \left(G^* - D x_\textnormal{couple} \right) \nonumber
\\
&& \textnormal{subject to} \quad  
\eqref{eq:PDN:BFM:power-balance-P}\text{--}\eqref{eq:PDN:BFM:voltage-drop}, \eqref{eq:PDN:voltage-limits}\text{--}\eqref{eq:PDN:power-injection-q}, \eqref{eq:PDN:BFM:soc}
\nonumber
\end{eqnarray}  
where constant matrix $D$ stacks $(D_\textnormal{water}, D_\textnormal{heat})$ and vectors $\mu^*$ and $G^*$ are received from solving appropriate OWF-C and OHF-C, which will be explained later. The OPF-C problem is thus parameterized as $\textnormal{OPF-C}(\mu^*, G^*)$.  
\begin{eqnarray}
\textbf{OWF-C}: && \min_{\tilde x_\textnormal{water}} \quad  \lambda_{\textnormal{water}}^t  q_{\textnormal{W0}}^t
\nonumber
 \\
 && \textnormal{subject to}  \quad 
 \eqref{eq:MWN:pipe-limit}\text{--}\eqref{eq:MWN:pump-head-gain}, \eqref{eq:MWN:valve-head-drop} 
 \nonumber
 \\
 && \quad G_\textnormal{water}(\tilde x_\textnormal{water}^o; \tilde x_\textnormal{water}) \leq D_\textnormal{water} x_\textnormal{couple}^* ~:\mu_{\textnormal{water}} \nonumber
\end{eqnarray}  
which, with a trivial objective, is a feasibility problem given $x_\textnormal{couple}^*$ received from solving an appropriate OPF-C. Vector $\tilde x_\textnormal{water}$ augments $x_\textnormal{water}$ by auxiliary variables $(W, \gamma, z)$ that were introduced to convexify \eqref{eq:MWN:pump-power}, \eqref{eq:MWN:Darcy-Weisbach}. The convexified constraints \eqref{eq:Darcy-Weisbach:soc}, \eqref{eq:MWN:pump-power:linear}, \eqref{eq:MWN:pump-power:convex-1}, \eqref{eq:MWN:pump-power:convex-2} are represented compactly by the last inequality in OWF-C, where constant matrix $D_\textnormal{water}$ extracts from $x_\textnormal{couple}^*$ appropriate expressions of $p^{\textnormal{w-pump},*}$; function $G_\textnormal{water}(\tilde x_\textnormal{water}^o; \cdot)$ is convex in $ \tilde x_\textnormal{water}$ and is parameterized by reference point $\tilde x_\textnormal{water}^o$ for convex-concave procedure; $\mu_{\textnormal{water}}$ is the associated vector of dual variables. The OWF-C problem can thus be denoted as $\textnormal{OWF-C}(x_\textnormal{couple}^*, \tilde x_\textnormal{water}^o)$.
\begin{eqnarray}
\textbf{OHF-C}: && \min_{\tilde x_\textnormal{heat}} \quad \sum_{t \in \T}  \sum_{i \in \N^{\textnormal{H,G}}}  \sum_{k \in \K_i} C_{i,k} \alpha^t_{i,k}   
\nonumber
 \\
 && \textnormal{subject to}  \quad 
 \eqref{eq:CHP:heatgen}\text{--}\eqref{eq:CHP:convex-2}, \eqref{eq:CHP:supplyTbound}\text{--}\eqref{eq:CHP:pump-head-gain}, \eqref{eq:heatload:returnTbound}\text{--}\eqref{eq:DHN:heatjunction:head}
 \nonumber
 \\&& \qquad\qquad\quad
 \eqref{eq:DHN:heatpipe-limit:S}, \eqref{eq:DHN:heatpipe-limit:R}, \eqref{eq:heatnode:continuity:S}, \eqref{eq:heatnode:continuity:R}
 \nonumber
 \\ && ~
 G_\textnormal{heat}(\tilde x_\textnormal{heat}^o; \tilde x_\textnormal{heat}) \leq D_\textnormal{heat} x_\textnormal{couple}^* ~:\mu_{\textnormal{heat}} \nonumber
\end{eqnarray}  
where $x_\textnormal{couple}^*$ is received from solving an appropriate OPF-C. Vector $\tilde x_\textnormal{heat}$ augments $x_\textnormal{heat}$ by auxiliary variables $(W, \gamma, z, y)$ that were introduced to convexify \eqref{eq:CHP:heating}, \eqref{eq:CHP:pump-power}, \eqref{eq:heatload:heating}, \eqref{eq:DHN:Darcy-Weisbach:heatpipe:S}, \eqref{eq:DHN:Darcy-Weisbach:heatpipe:R}, \eqref{eq:thermal-propagation:S}--\eqref{eq:heatnode:temperature-mix:R}. The last inequality in OHF-C compactly represents convexified constraints such as \eqref{eq:Darcy-Weisbach:soc}, \eqref{eq:MWN:pump-power:convex-1}, \eqref{eq:MWN:pump-power:convex-2}, \eqref{eq:thermal-propagation:log:convex}, \eqref{eq:thermal-propagation:log:linearize} and coupling constraints \eqref{eq:CHP:powergen:p}, \eqref{eq:CHP:powergen:q}. In particular, constant matrix $D_\textnormal{heat}$ extracts from $x_\textnormal{couple}^*$ appropriate expressions of $(p^{\textnormal{gen},*}, q^{\textnormal{gen},*}, p^{\textnormal{h-pump},*})$; function $G_\textnormal{heat}(\tilde x_\textnormal{heat}^o; \cdot)$ is convex in $ \tilde x_\textnormal{heat}$ and is parameterized by reference point $\tilde x_\textnormal{heat}^o$ for convex-concave procedure; $\mu_{\textnormal{heat}}$ collects dual variables. The OHF-C problem can thus be denoted as $\textnormal{OHF-C}(x_\textnormal{couple}^*, \tilde x_\textnormal{heat}^o)$.
 
Based on the decomposition above, we propose Algorithm \ref{alg:decomposed} to solve OPWHF. A flowchart of Algorithm \ref{alg:decomposed} is provided in Figure \ref{fig:flowchart}.
What is done by Algorithm \ref{alg:decomposed} is self-explanatory and the underlying idea was explained at the beginning of this section. We just add three remarks.

First, the convergence criterion in Algorithm \ref{alg:decomposed} may be a maximum number of iterations or sufficiently small changes in decision variables and/or objective values.

Second, to find an initial feasible point for Algorithm \ref{alg:decomposed}, we formulate a feasibility problem with nonnegative slack variables, and then minimize an increasing function of those slack variables to make them as close to zero as possible. 
The iterative feasible point pursuit algorithm in \cite{zamzam2018optimal} can solve such a nonlinear feasibility problem. In our experiment, we solve it by calling the SCIP solver from OPTI (https://www.inverseproblem.co.nz/OPTI/index.php). 

Third, the proposed method utilizes the convex-concave procedure, which only guarantees a feasible local optimum, as proved in reference \cite{lipp2016variations}. Seeking and proving for the global optimum of OPWHF seems much harder and requires approaches such as exact convex relaxation.

\section{Numerical Results}\label{sec:simulation}

We simulate the system in Figure \ref{fig:1}, which comprises a single-phase version of the IEEE 37-node power distribution network, a water network of 5 nodes, and a district heating network of 5 nodes. 
The parameters of the IEEE 37-node network are from \cite{IEEE37system}; the parameters of the water network, such as the pump efficiency ($\eta = 0.81$), the initial inlet head at the tank ($5\textnormal{m}$), and the head at the reservoir ($25\textnormal{m}$), are from \cite{fooladivanda2017energy}; the parameters of the district heating network are from \cite{li2016DHoptLinear}, \cite{li2016DHoptNonlinear}, \cite{fooladivanda2017energy}, \cite{lahdelma2003CHP}. See our online data file for detail.\footnote{https://gocuhk-my.sharepoint.com/:x:/g/personal/chzhao\_cuhk\_edu\_hk /EcE5faFL-9ZMjPXCOm9CT2QBUd8xJL1YiFlQQnPgPrFKDA}

\begin{figure}
	\includegraphics[width=1.0\columnwidth]{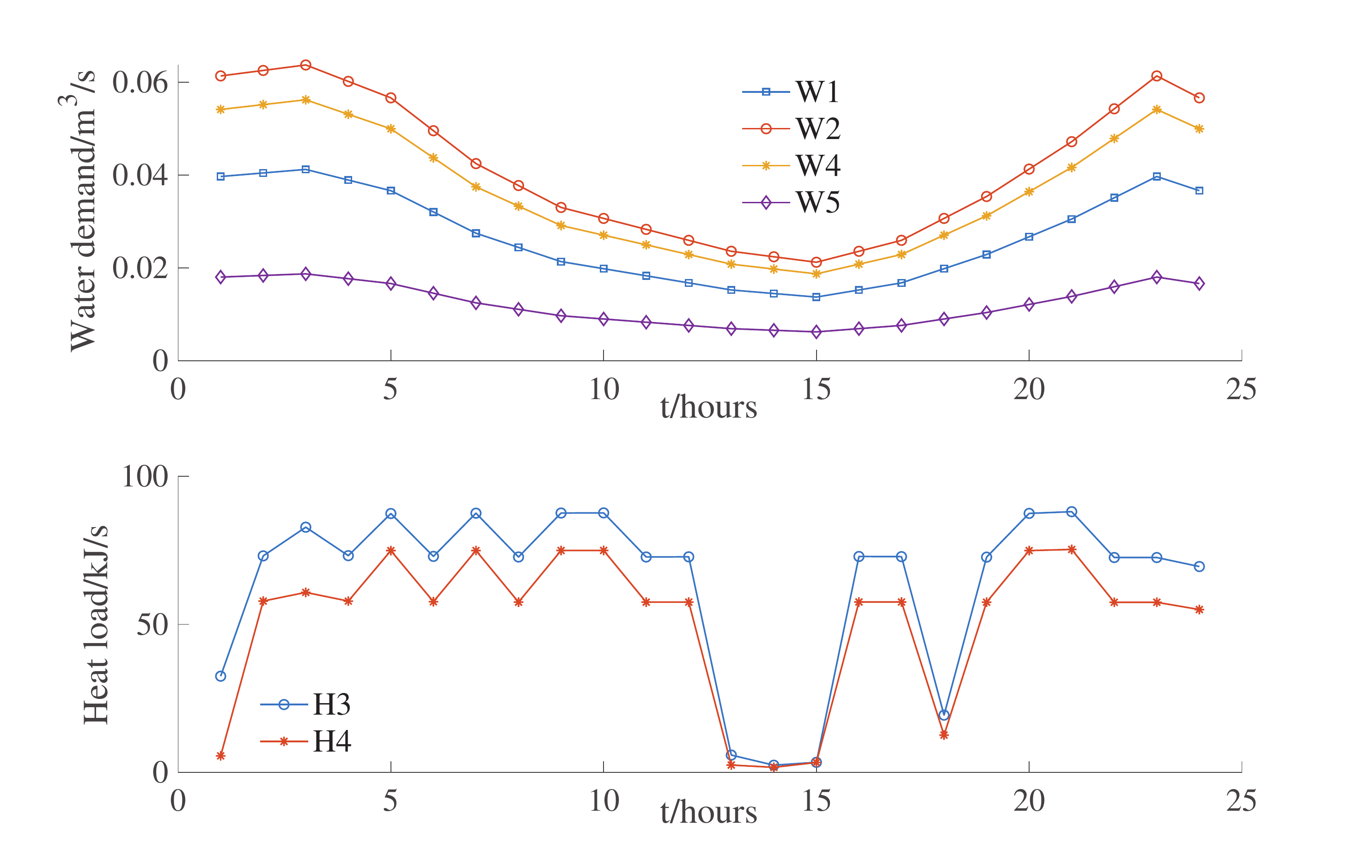}
	\centering
	\caption{Demands for water (upper) and heat (lower) in the simulation.}
	\label{fig:demand}
\end{figure}

\begin{figure}
	\includegraphics[width=0.7\columnwidth]{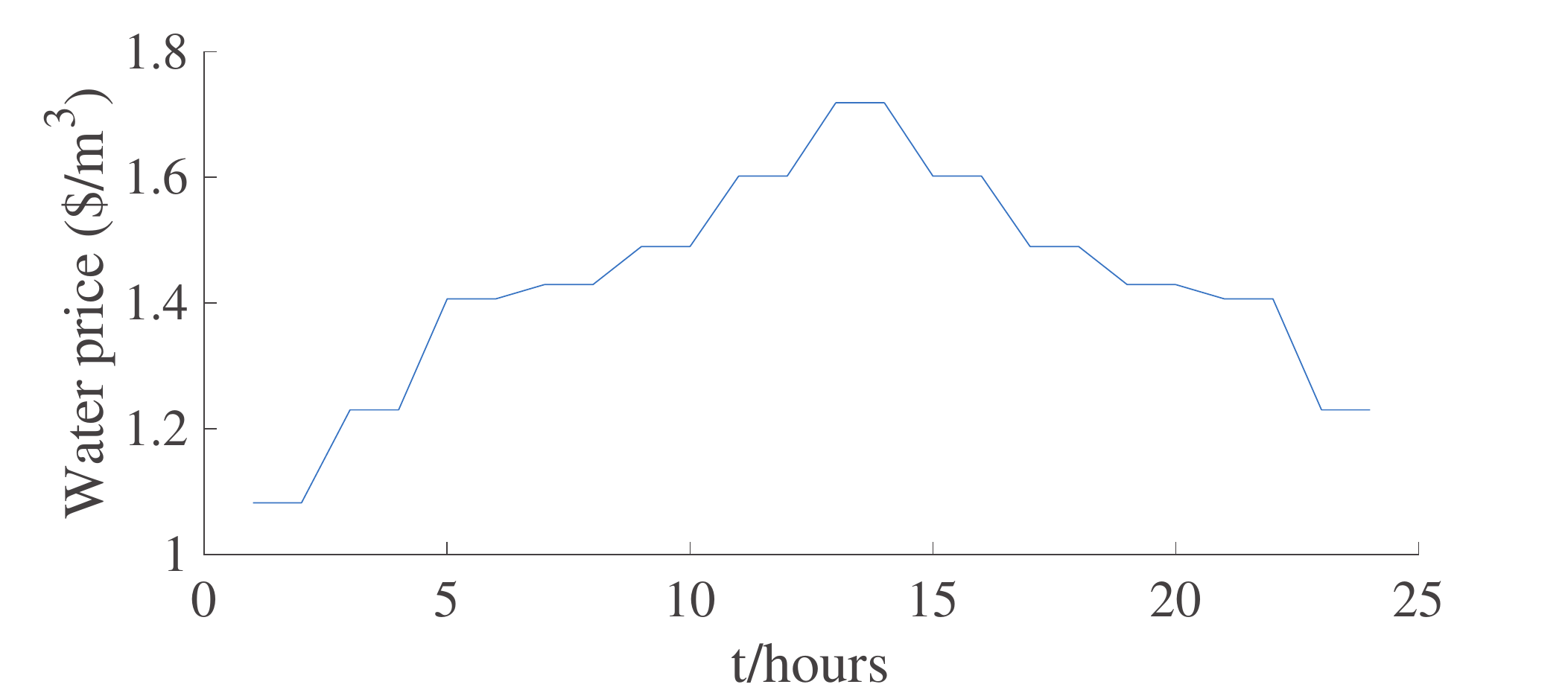}
	\centering
	\caption{Water price used in the simulation.}
	\label{fig:waterprice}
\end{figure}

\begin{figure}
	\includegraphics[width=1.0\columnwidth]{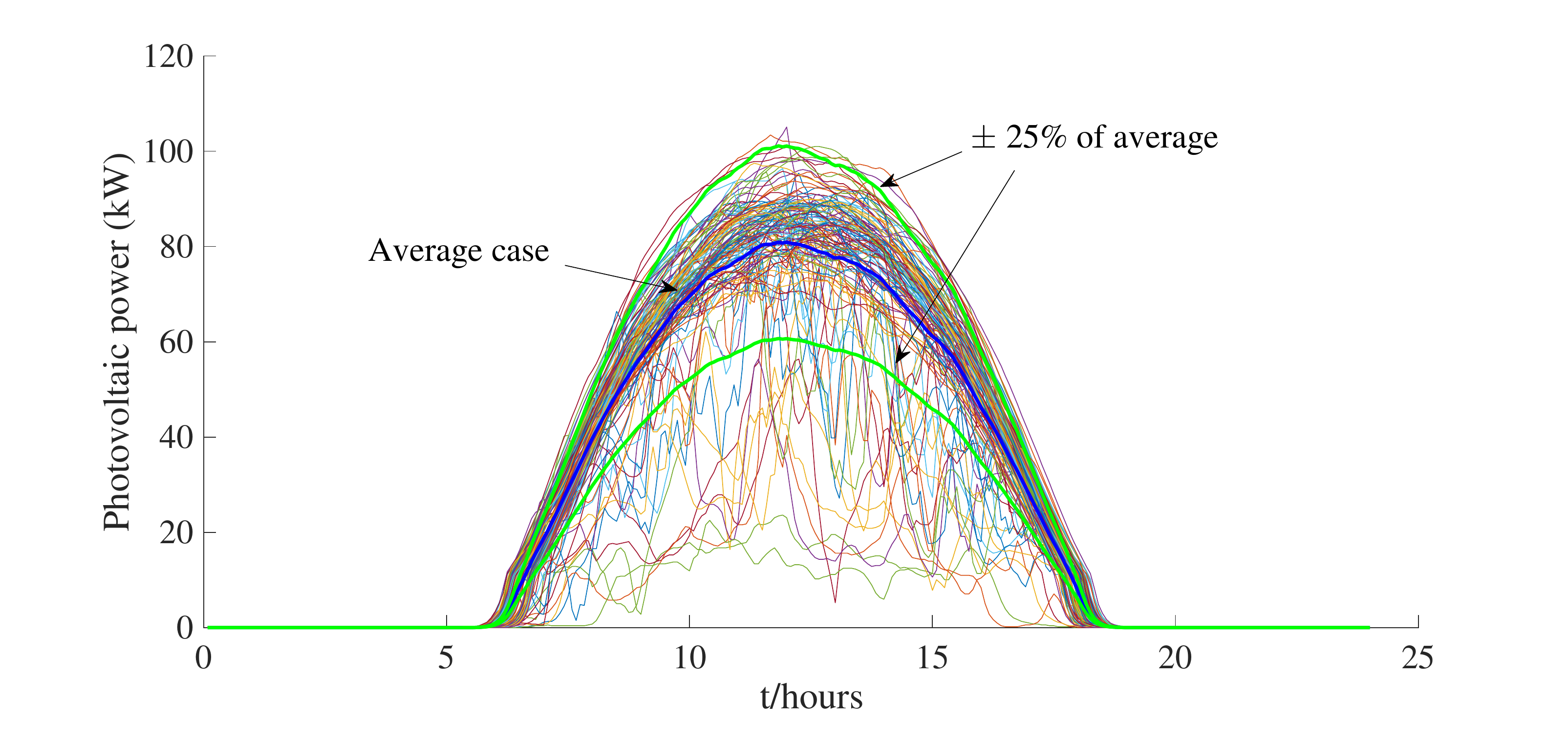}
	\centering
	\caption{The 5-minute power output of a PV system in California, over 123 summer days. Data source: NREL \cite{PVdataset}.} 
	\label{fig: PVpowerCalifornia}
\end{figure}

The water and heating demands for our simulation are shown in Figure \ref{fig:demand}.
We adopt the electricity price profile (Figure \ref{fig: DRW}(a), Case D) of the Midcontinent Independent System Operator on June 20, 2017; the same price was used in \cite{zamzam2018optimal}.
The water price used in the simulation is provided in Figure \ref{fig:waterprice}.
In some of our test scenarios, we integrate a solar photovoltaic (PV) system at node E22 of the power network. Figure \ref{fig: PVpowerCalifornia} displays the 5-minute generation data for a PV system in California, over 123 days from May to August 2006; the data set is provided by NREL \cite{PVdataset}, with PV capacity scaled down to fit the size of our system. For the scheduling problem concerned, we take hourly samples of the average of the 123 days (the average curve in Figure \ref{fig: PVpowerCalifornia} is sampled as the ``original'' case in Figure \ref{fig: PVpower}) and treat the obtained profile as the maximum available PV power over a typical summer day.

\subsection{Results of joint optimization}
\label{subsec:simulation:performance}

\subsubsection{Convergence of the proposed algorithm}

\begin{figure}
	\includegraphics[width=0.7\columnwidth]{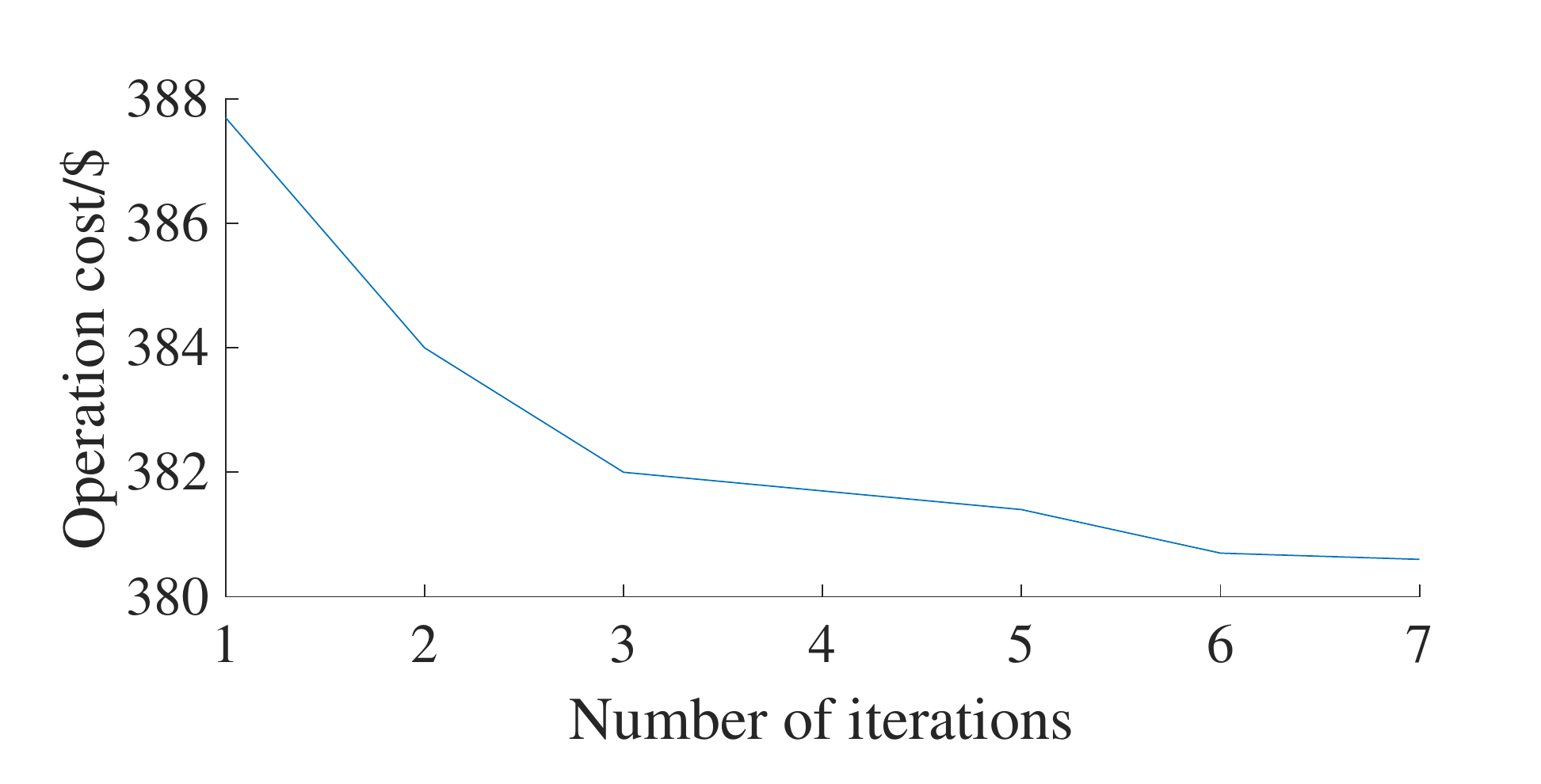}
	\centering
	\caption{Convergence of the OPWHF objective value.}
	\label{fig: convergence}
\end{figure}

Figure \ref{fig: convergence} shows that the OPWHF objective value converges in seven iterations, which takes $573.4$ seconds on a PC with Intel Quad-Core i7 CPU and 16GB RAM, using Gurobi 9.0 called from MATLAB CVX as the optimization solver.

\subsubsection{Cost reduction by joint optimization}   

Table \ref{Table: result comparison} compares costs between the separate and the proposed joint optimization schemes. The costs to three networks are all reduced by the joint optimization, due to the flexibility added by integrating multiple networks. 
As a remark, the cost to water network refers to the water plus electricity payment.  
The cost to heating network refers to the CHP generation cost plus the electricity payment for the CHP water pumps, minus the CHP electricity generation revenue. The cost to power network is offset by the electricity payment and revenue of the heating and water networks to avoid double counting. 

\begin{table}\centering
	\caption{Costs under separate and joint optimizations}\label{Table: result comparison}
	\begin{tabular}{c|ccc|c}
		\hline
		Costs ($\$$) &  Power & Water & Heating  &Total           \\ 
		\hline
		Separate & $330.4$   & $18.5$ & $38.9$ &$387.8$\\
		Joint       & $328.7$   & $13.4$ & $38.5$ & $380.6$\\ 
		\hline
	\end{tabular}
\end{table}

\subsubsection{Improved PV utilization by joint optimization}  

\begin{figure}
	\includegraphics[width=0.9\columnwidth]{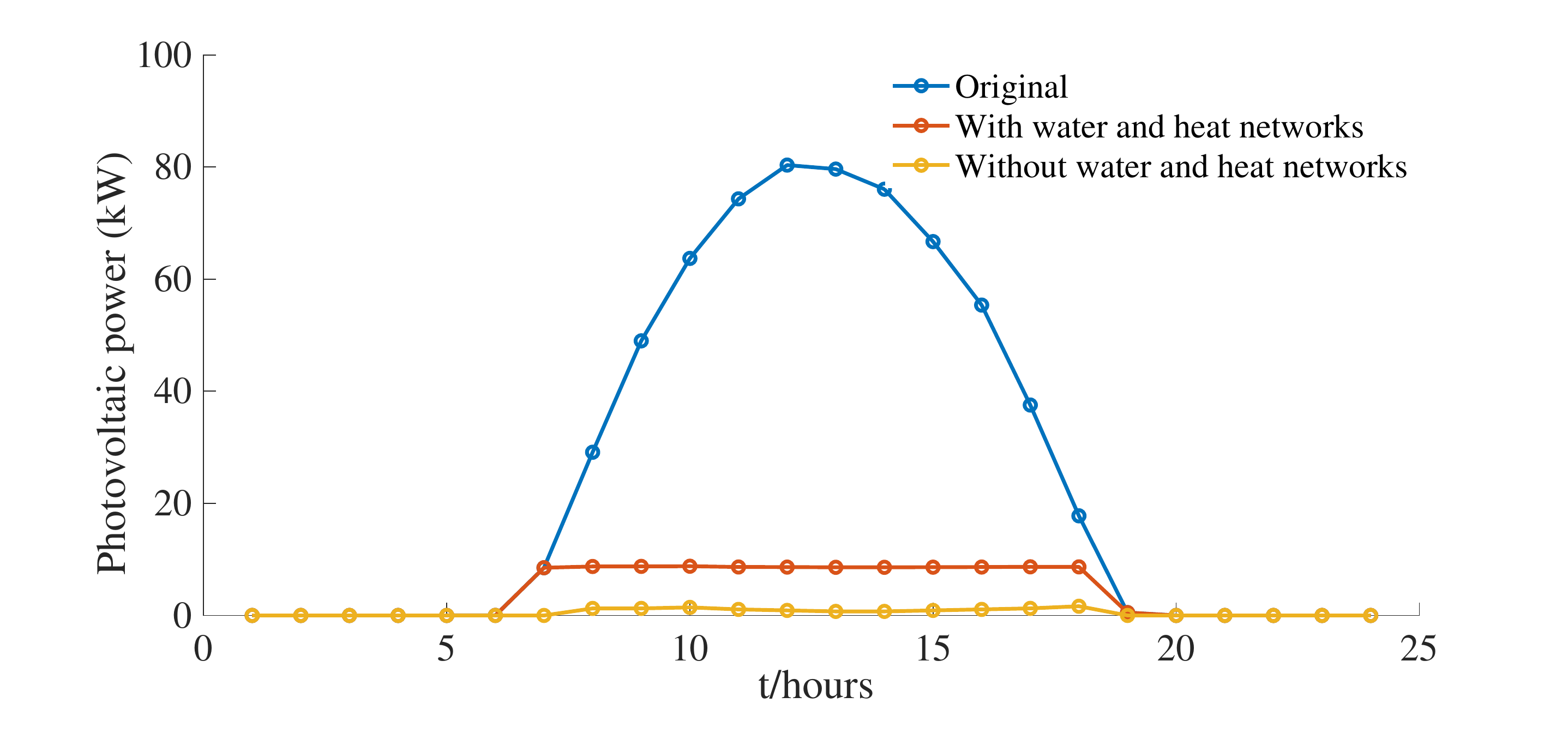}
	\centering
	\caption{PV power utilization with and without water and heating network integration. The ``original'' refers to the maximum available PV power.}
	\label{fig: PVpower}
\end{figure} 

The utilized PV power outputs with and without water and heating network integration are plotted in Figure \ref{fig: PVpower} and compared with the maximum available PV power. We observe a higher utilization of PV power through integrating water and heating networks, mainly due to the water pumps that use excessive PV power to pump water into the pipes and tanks as virtual storage. Otherwise, the excessive PV generation would just be curtailed and wasted.  

\begin{figure}
	\includegraphics[width=1.0\columnwidth]{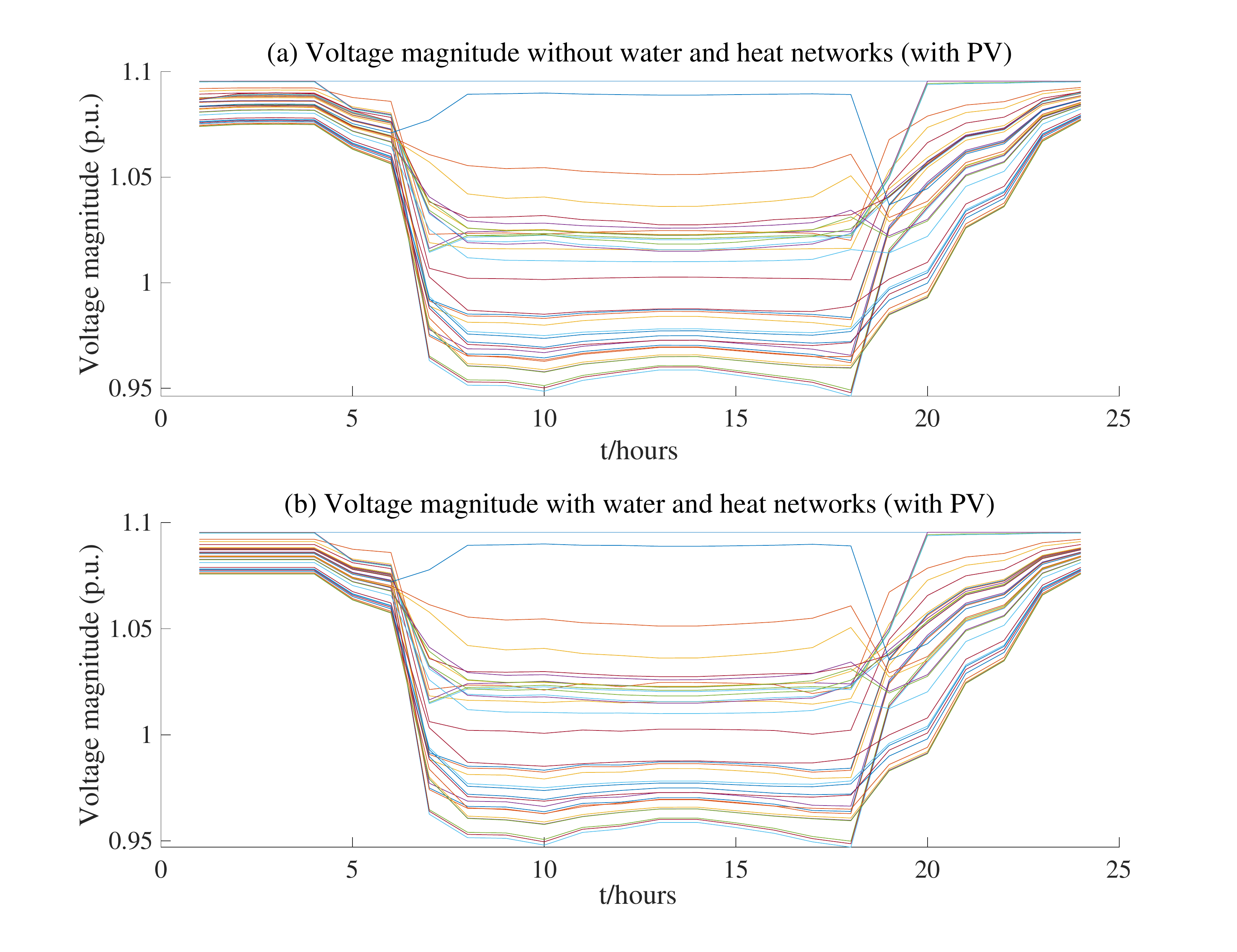}
	\centering
	\caption{Voltage versus time without (upper) and with (lower) water and heating network integration. Each curve corresponds to a power network node.}
	\label{fig: voltageprofile}
\end{figure}

Figure \ref{fig: voltageprofile} shows voltage magnitudes versus time at different power network nodes with/without the integration of water and heating networks. We observe that this integration does not adversely affect reliability and security of the power network, in the sense that the voltage magnitudes are all maintained inside the safety region $[0.95,~1.1]$ per unit. Indeed, the voltage changes caused by integrating water and heating networks are not significant, due to the capability of power network to adjust DER power outputs for voltage regulation \eqref{eq:PDN:voltage-limits}.

  \subsubsection{Influence of uncertainties}  
  Different patterns of solar generation and demand in different seasons can lead to different levels of uncertainties in generation and load forecasts. In \cite{golestaneh2016very}, the short-term solar forecast errors in four seasons are respectively 8.08\%, 10.73\%, 8.48\%, 4.16\%. We thus experiment with the standard deviation of PV power ranging in 0--10\% of its expectation (i.e., the original case in Figure \ref{fig: PVpower}). 
The maximum standard deviation $\sigma=10\%$ also fits the observation in Figure \ref{fig: PVpowerCalifornia} that most of the actual PV outputs lie in the $\pm 2.5\sigma = \pm 25\%$ interval.
  In \cite{WILLIAMS2020178}, the average forecast error for residential demand is 4.55\%, and we accordingly set the standard deviations of power, water, and heat demands at 0--5\% of their expectations (i.e., the original electric loads in the IEEE 37-node network and water and heating demands in Figure \ref{fig:demand}). The consequent deviations of the OPWHF objective value from its expectation are 
  estimated using the three-point method \cite{su2005probabilistic} and shown in Table \ref{Table: uncertainties}. It is observed that higher uncertainties measured by larger standard deviations can lead to a moderate increase of deviation in the optimal value.
  
  \begin{table}\centering
  	\caption{Influence of uncertainties on OPWHF objective value}\label{Table: uncertainties}
     \begin{tabular}{cccc}
	\hline
	& \multicolumn{3}{c}{Standard deviation} \\ \hline
	Solar generation       & 0\%        & 5\%         & 10\%        \\
	Electric demand  & 0\%        & 2.5\%       & 5\%         \\
	Water demand      & 0\%        & 2.5\%       & 5\%         \\
	Heating demand     & 0\%        & 2.5\%       & 5\%         \\
	Objective value & 0\%        & 3.4\%       & 6.8\%       \\ \hline
    \end{tabular}
  \end{table}

\subsection{Characterizing added flexibility by joint optimization}\label{subsec:added_flexibility}

\subsubsection{Flexibility to utilize PV} 
We first study the improvement in network flexibility by the joint optimization, which leads to better utilization of PV generation.

\begin{figure}
	\includegraphics[width=0.9\columnwidth]{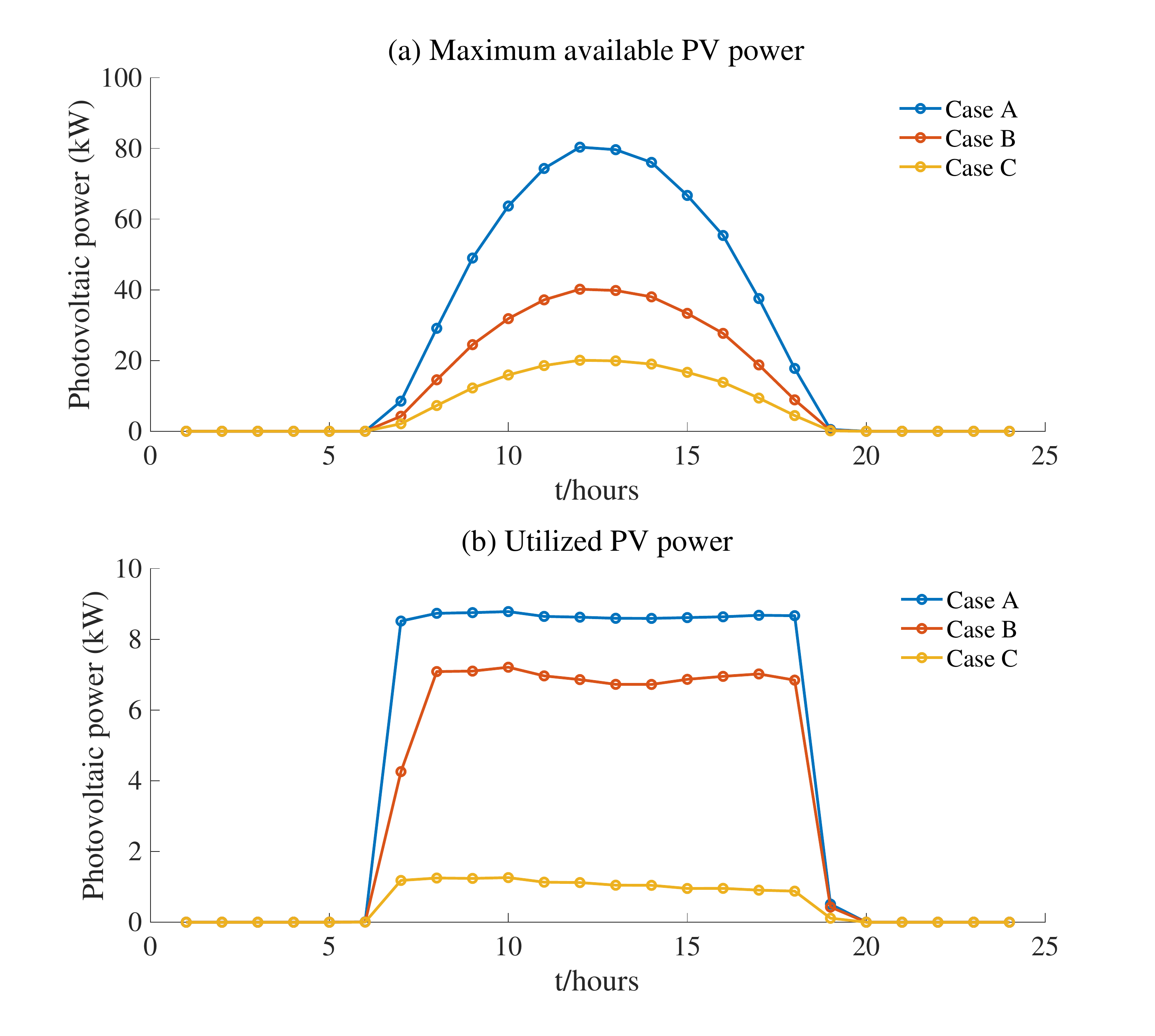}
	\centering
	\caption{The PV power utilized with water and heating network integration, for three different cases of maximum available PV power.}
	\label{fig: improvePVintegration}
\end{figure}

Consider three scenarios of maximum available PV power as shown by Cases A--C in Figure \ref{fig: improvePVintegration}(a). 
The corresponding PV power utilized by the integrated network (including water and heat) is plotted in Figure \ref{fig: improvePVintegration}(b). We observe that the utilized PV power increases with the maximum available PV power. This observation reveals the  potential of the proposed joint optimization framework in providing higher flexibility to integrated energy systems and saving more cost under a higher proportion of variable renewable generation. 

\subsubsection{Flexibility to cope with price variations}

\begin{figure}
	\includegraphics[width=1.0\columnwidth]{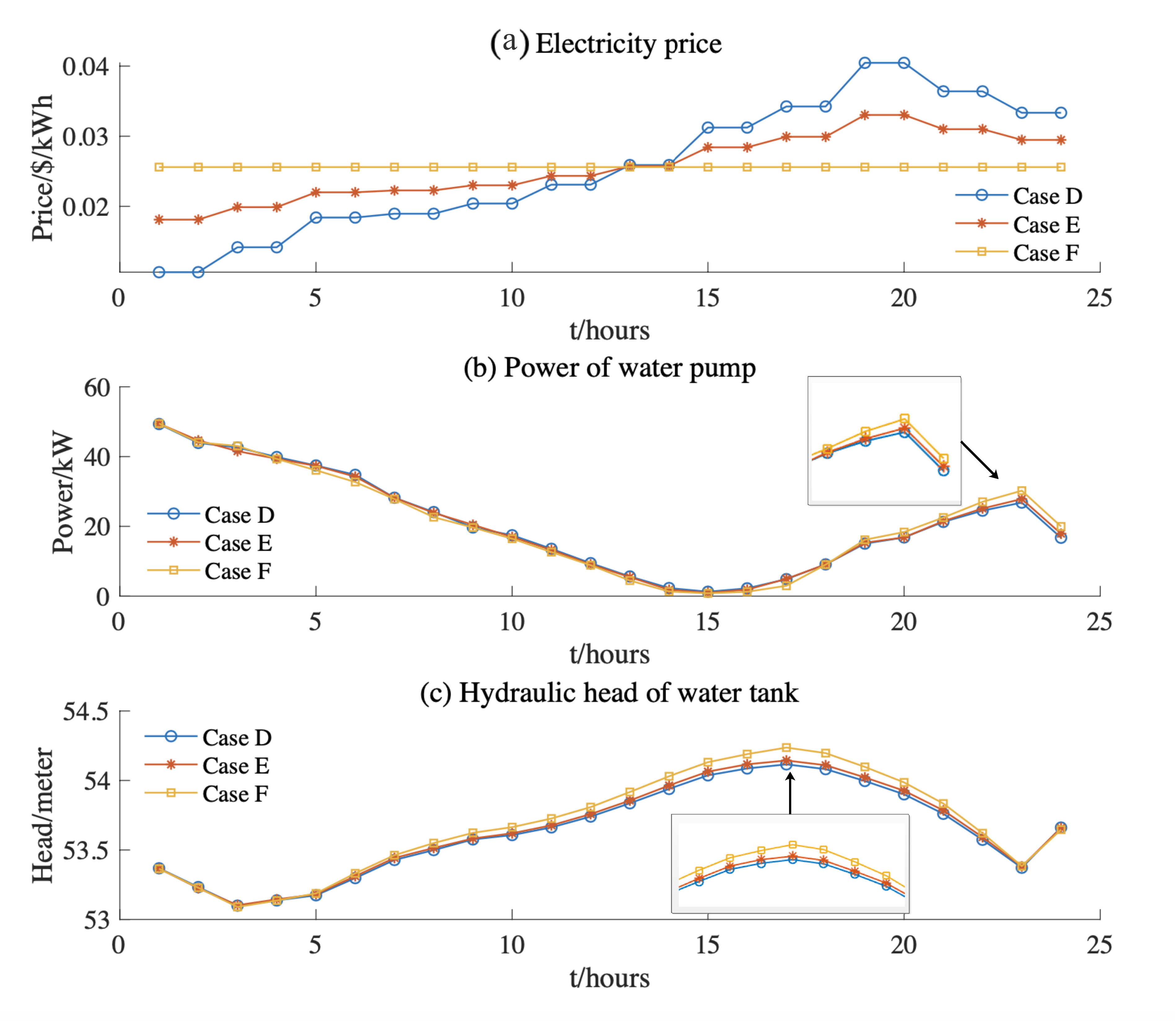}
	\centering
	\caption{Response of water network to different electricity price profiles.}
	\label{fig: DRW}
\end{figure}

Consider three electricity price profiles as Cases D--F in Figure \ref{fig: DRW}(a), to which the corresponding power of water pump and head of water tank are shown in Figures \ref{fig: DRW}(b) and \ref{fig: DRW}(c), respectively. The general trend is that the tank water head is elevated, i.e., the tank is charged, at low price, and discharged as price grows. Under larger price variation, the pump tends to consume more power at low price and less at high price. More energy is exchanged to respond to larger price variation, and thus less energy in the form of water head is stored, as indicated by the lower peak in Figure \ref{fig: DRW}(c). These results demonstrate the role of the water tank as a virtual energy storage to enhance system flexibility. 

\begin{figure}
	\includegraphics[width=1.0\columnwidth]{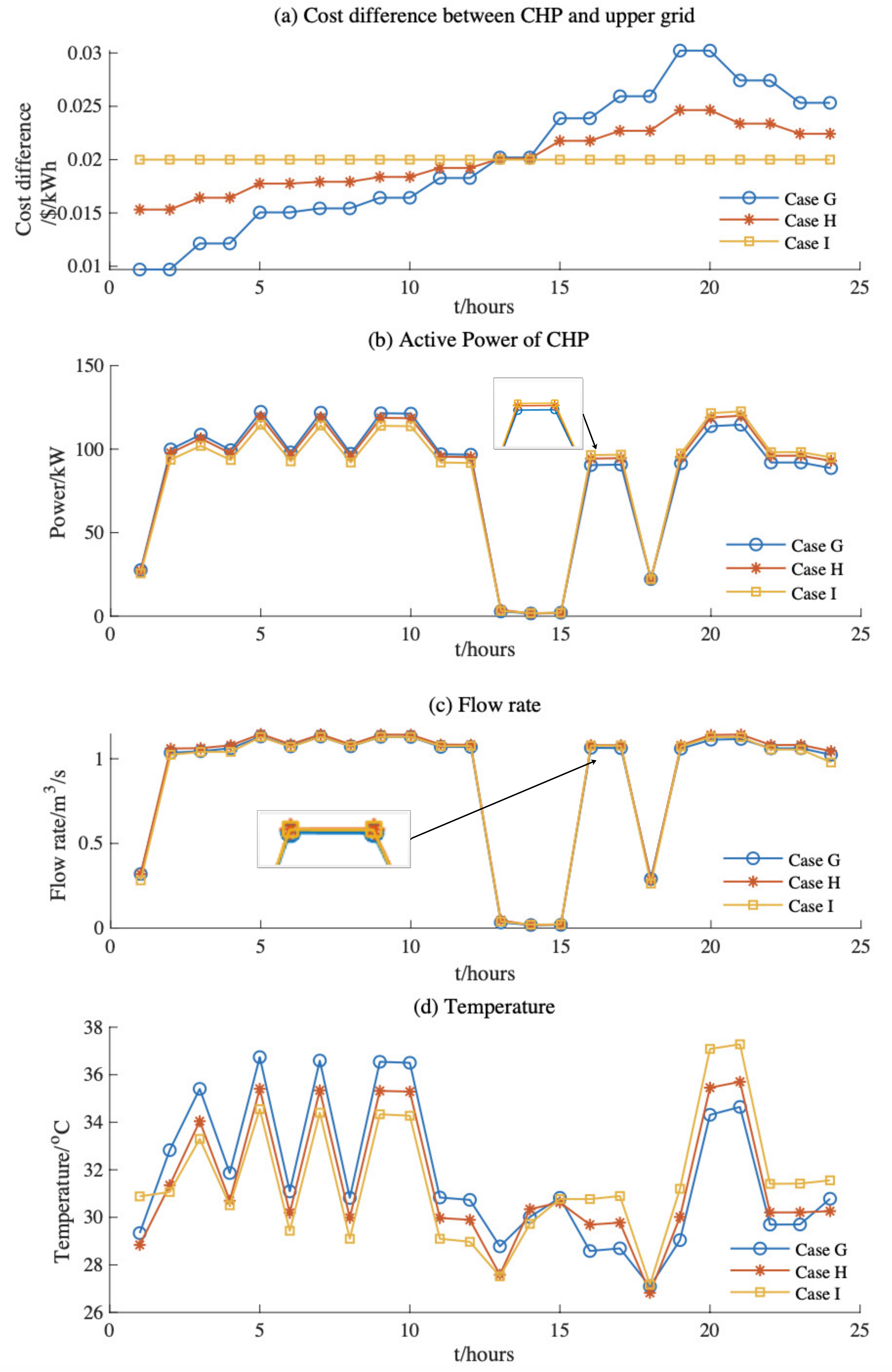}
	\centering
	\caption{Response of heating network to different CHP generation costs compared to the electricity price from the upper grid.}
	\label{fig: DRH}
\end{figure}

We then consider three scenarios of cost difference between CHP and the main grid, as Cases G--I in Figure \ref{fig: DRH}(a), where a larger difference means CHP generation is more expensive than importing electricity from the main grid. The corresponding active power generation, pump flow rate, and temperature enhancement of the CHP are shown in Figures \ref{fig: DRH}(b)--(d), respectively. We observe that the CHP generates more power and raises to a higher temperature at relatively lower generation cost, and vice versa, while the difference in pump flow rate is not that significant. In this sense, the heating network also serves as an energy storage to add system flexibility, by utilizing its thermal inertia.  

\subsection{Sensitivity analysis for heating network control} \label{subsec:sensitivity}

\begin{figure}
	\includegraphics[width=1.0\columnwidth]{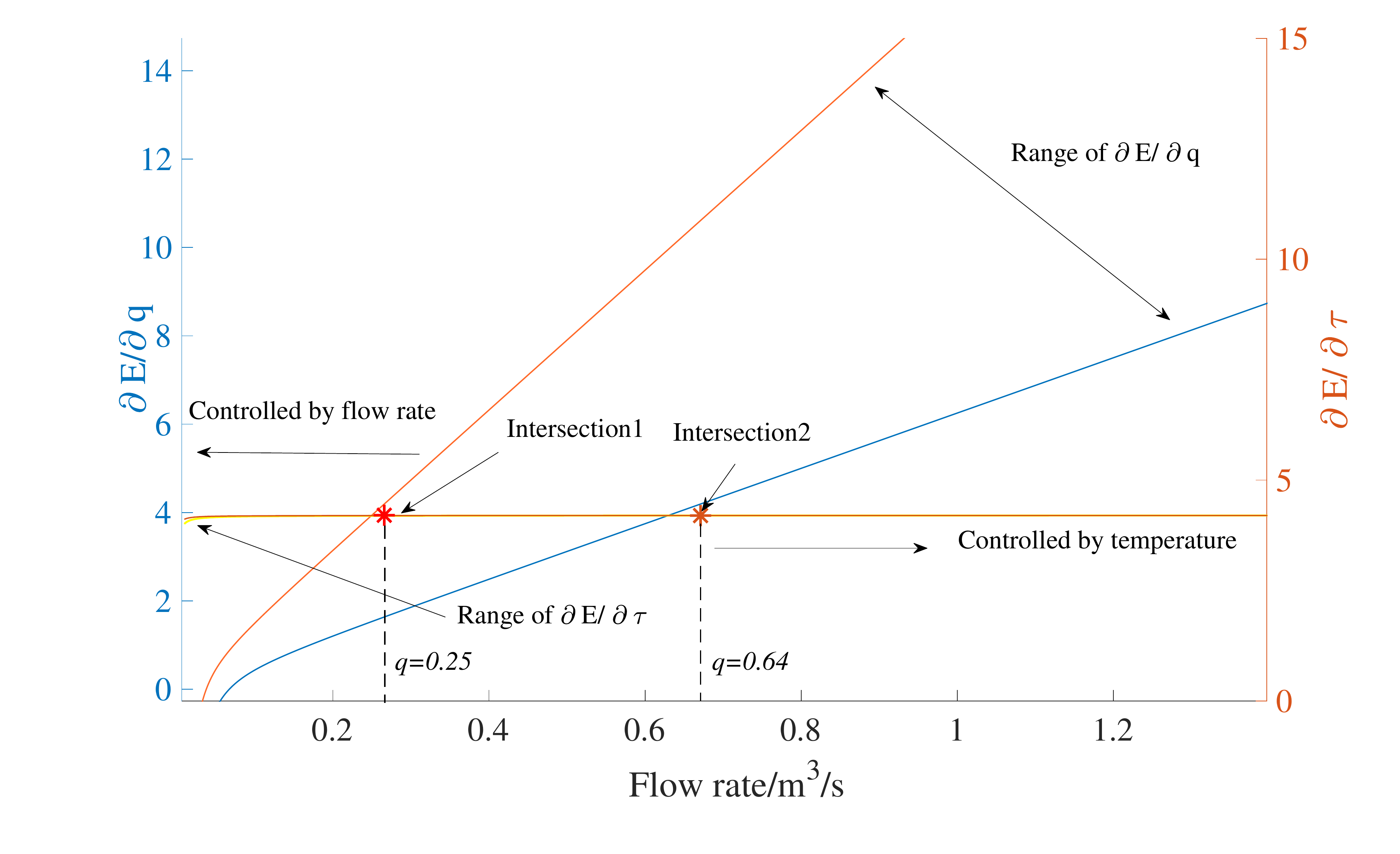}
	\centering
	\caption{Relationship between $\frac{\partial E}{\partial q}$ and $\frac{\partial E}{\partial \tau}$ to decide preferred control mode.}
	\label{fig: ORange}
\end{figure}

There are two basic control modes of the heating network, i.e., the flow-rate control mode  which adjusts flow rates $q$ and the temperature control mode which adjusts the temperature $\tau$. 
%The mixed mode of controlling both $\tau$ and $q$ can be regarded as a combination of the two basic modes. 
We next compare effectiveness of the two basic modes, as well as their mixture, in different heating networks characterized by their pipe parameters (lengths, diameters, and friction factors).
This comparison is conducted by analyzing sensitivities of energy loss $E = mg \Delta h + C m \Delta \tau = mg F q^2  + Cm (\tau - \tau_0) e^{-\frac{\xi}{q}}$ for given mass $m$ of water along a pipe. Partial derivatives $\partial E/\partial \tau$ and $\partial E/\partial q$ can change across flow rate $q$ within a range of pipe parameters, as shown in Figure \ref{fig: ORange} (where the two boundary lines for $\partial E/\partial \tau$ are very close). The preferred control mode is the one with lower sensitivity to reduce energy loss. Therefore, to the left of Intersection 1, it is preferred to control by $q$; to the right of Intersection 2, it is preferred to control by $\tau$, and in area between Intersection 1 and 2, it is preferred to control by both.

\begin{figure}
	\includegraphics[width=0.8\columnwidth]{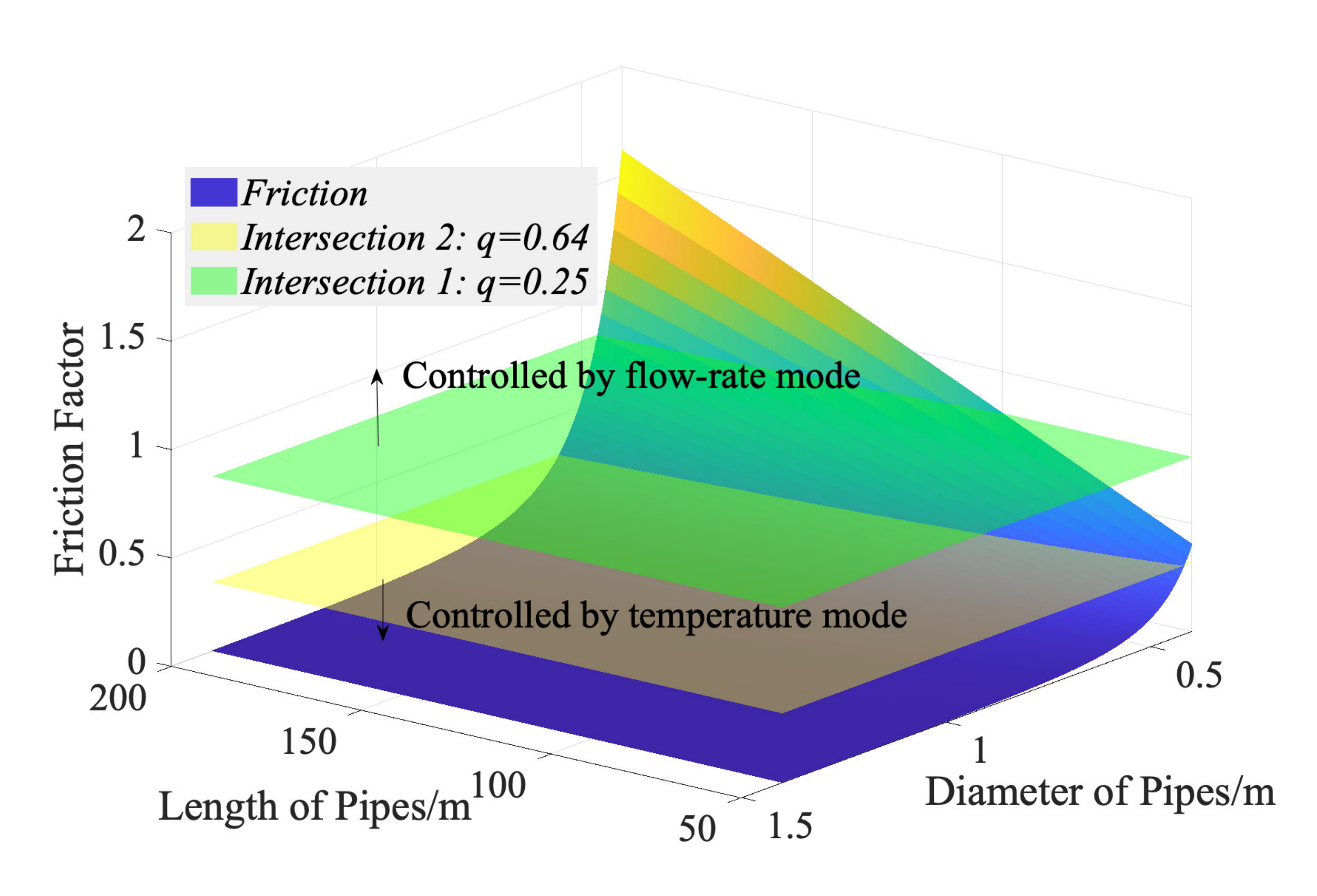}
	\centering
	\caption{Control modes of heating networks with different pipe parameters.}
	\label{fig: 3Drange}
\end{figure}

Two critical friction factors can be solved from Intersections 1 and 2 and plotted in Figure \ref{fig: 3Drange} together with the surface that defines the relationship between friction factor and pipe diameter and length. The surface is thus divided into three parts with different preferred control modes. 
In general, controlling by temperature is more effective for most practical systems with ordinary pipe parameters since it covers the most part of the parameter space, while controlling by flow rate is more effective to very long and thin pipes (longer than 150m, diameter smaller than 0.6m).

We discuss more on the role of the friction factor in effectiveness of the two control modes. As shown in Figure \ref{fig: 3Drange}, the temperature control mode is more effective in the parameter space with a low friction factor. Only when the friction factor becomes larger than a certain threshold the flow-rate control mode becomes more effective.  
Note that the friction factor $F$ enters the expression of $\partial E/\partial q$ and thus affects the relationship between $\partial E/\partial q$ and $\partial E/\partial \tau$ as shown in Figure \ref{fig: ORange}, to decide whether it is more effective to control flow rate $q$ or temperature $\tau$.

\section{Conclusion}\label{sec:conclusion}
We formulated an optimal power-water-heat flow (OPWHF) problem and solved it via a decomposed iterative convex approximation heuristic. Numerical results validated convergence of the proposed method and improvement in social benefits and operational flexibility achieved by the joint optimal energy scheduling of multiple networks. 
Our main results are summarized as follows.
\begin{enumerate}
\item	The proposed joint optimization can add flexibility to the integrated system to improve its economic performance. Compared with separate operations of power, water, and heating networks, the total daily operational cost is reduced from \$387.8 to \$380.6, i.e., by 1.86\%; particularly, the cost to the water network is reduced from \$18.5 to \$13.4, i.e., by 27.57\%. 
\item	As the standard deviations of solar generation and demands (power, water, heat) increase from zero to 10\% and 5\% of their expectations, respectively, the standard deviation of the OPWHF objective value increases to 6.8\%, which reveals a moderate impact of those uncertainties on the proposed joint optimization framework. 
\item	The joint optimization can save significant electricity payment from the integrated system, by utilizing excessive PV generation to pump up the water tanks and adapting the total energy usage to time-varying prices. The water tanks and heating pipes can serve as energy storage to enhance system flexibility.
\item Sensitivity analysis and numerical results suggest that in the heating network, the temperature control mode is typically more effective than the flow-rate control mode, unless for networks with very long and thin pipes (longer than 150m, diameter smaller than 0.6m in our test cases). 
\end{enumerate}

We also notice some deficiencies of this work that call for improvement in the future:
\begin{enumerate}
\item	A simplified district heating network model remains to be developed to improve computational efficiency while preserving sufficient accuracy.
\item	An operational framework with tighter integration of power-gas, power-heat, fuel cell, etc. waits to be developed while considering realistic network models.
\item	We need to develop an integrated energy scheduling model which fully accounts for the uncertainties of renewable generation and demand, e.g., by utilizing stochastic or robust optimization methods.
\end{enumerate}

\section*{Acknowledgment}
We thank Kyri Baker, Emiliano Dall'Anese, Ahmed Zamzam, and the editor and reviewers for their valuable supports, comments, and suggestions that help improve our work. A preprint version of this article was posted on arXiv \cite{fang2021arXiv}.


\end{document}